\documentclass[
aps,
reprint,
prb,
superscriptaddress,
amsmath,
amssymb,
floatfix,
longbibliography,
]{revtex4-2}

 \UseRawInputEncoding

 \usepackage{amsmath,bm}
 \usepackage{mathrsfs}
 \usepackage{amsfonts}
 \usepackage{graphicx}
 \usepackage{setspace} 
 \usepackage{graphicx}
 \usepackage{epstopdf}
 \usepackage{dcolumn}
 \usepackage{amsmath}
 \usepackage{epsfig}
 \usepackage{indentfirst}
 \usepackage{psfrag}
 \usepackage{subfigure}
 \usepackage{amssymb}
 \usepackage{color}
 \usepackage{xcolor}
 \usepackage{siunitx}

 \usepackage{graphicx}
 \usepackage{dcolumn}
 \usepackage{bm}

\usepackage{physics}
\usepackage{dcolumn}
\usepackage{bm}
\usepackage[backref=none,bookmarksnumbered=true,bookmarks=true,bookmarksopen=true,colorlinks=true,
citecolor=blue,linkcolor=blue,anchorcolor=green,urlcolor=blue,unicode=false]{hyperref}

\usepackage{ulem}[normalem]

\makeatletter

\newcommand\colorsout[1]{\bgroup \markoverwith{\textcolor{#1}{\rule[0.5ex]{2pt}{0.4pt}}}\ULon}

\makeatother
\begin{document}
\title{ 
Superconducting Spin-Singlet QuBit in a Triangulene Spin Chain}
\author{Chen-How Huang}

\affiliation{Donostia International Physics Center (DIPC), 20018 Donostia--San Sebasti\'an, Spain}
\affiliation{Departamento de Pol{\'i}meros y Materiales Avanzados: Física, Qu\'{\i}mica y Tecnolog\'{\i}a, Facultad de Ciencias Qu\'{\i}micas,
Universidad del Pa\'{\i}s Vasco UPV/EHU, 20018 Donostia-San Sebasti\'an, Spain.}
\affiliation{Department of Physics and Nanoscience Center, University of Jyväskylä, P.O. Box 35 (YFL), FI-40014 University of Jyväskylä, Finland}

\author{Jon Ortuzar}
\affiliation{CIC nanoGUNE-BRTA, 20018 Donostia-San Sebasti\'an, Spain }
\affiliation{Quantronics group, Service de Physique de l'\'Etat Condens\'e \mbox{(CNRS, UMR 3680)}, IRAMIS, CEA-Saclay, Universit\'e Paris-Saclay, 91191 Gif-sur-Yvette, France}
\newcommand{\eqContrib}{\thanks{These authors contributed equally to this work.}}
\author{M. A. Cazalilla}
\affiliation{Donostia International Physics Center (DIPC), 20018 Donostia--San Sebasti\'an, Spain}
\affiliation{IKERBASQUE, Basque Foundation for Science, Plaza Euskadi 5
48009 Bilbao, Spain}

\begin{abstract}
Chains of triangular nanographene (triangulene), recently identified as realizing  the valence-bond solid phase of a spin-$1$ chain, offer a promising platform for quantum information processing. We propose a  spin-singlet qubit based on these chains grown on a superconducting substrate. Using the numerical renormalization group (NRG), we identify a manifold consisting of the two lowest-lying, spin-singlet states isolated from doublet states of opposite fermion parity, which undergo an avoided crossing. A qubit utilizing these states is thus protected from random Zeeman and/or spin-orbit coupling. Despite the unavoidable effect of quasi-particle poisoning on qubit performance, the isolation of the singlet states offers additional protection. In addition, we introduce a mesoscopic device architecture, based on a triple quantum dot coupled to a superconducting junction, that quantum simulates the spin chain and enables control and readout of the qubit.  An effective two-level description of the device is validated using time-dependent NRG.
\end{abstract}
\date{\today}
\maketitle

\section{Introduction}

Chains of magnetic atoms and molecules have attracted considerable interest due to their potential to host exotic excitations and phases~\cite{Giamarchi2003,Haldane1983,HALDANE1983464,PhysRevLett.59.799,Affleck1990}.  Their low dimensionality enhances strong electron correlation effects~\cite{Giamarchi2003}, resulting in rich quantum behavior and offering promising platforms for realizing topological phases of matter~ \cite{PhysRevLett.59.799,Affleck1990,Kitaev2001,Pollmann2012}.  Indeed, fabricating magnetic chains on superconductors may enable the engineering of decoherence-free quantum memories~\cite{Kitaev2001} in topological superconductors hosting Majorana bound states~\cite{Bernevig2013,Yazdani2014}.

Building on recent advances, the on-surface synthesis technique~\cite{grill2007,gourdon2008} allows for precise engineering of carbon-based molecules and the manipulation of their magnetic properties. The spin state of several types of nanographene structures can be controlled in this way~\cite{pavlicek2017,Mishra2019,vilas2023surface}. Recently, the authors of  Ref.~\onlinecite{mishra2021} have demonstrated that a spin chain can be grown from triangulenes (i.e.,triangular nanographene structures). The triangulene spin chains (TSCs) were shown to realize the valence-bond solid (VBS) phase of a spin-$1$ chain~\cite{Haldane1983,HALDANE1983464,PhysRevLett.59.799,Affleck1990}, with open TSCs hosting topological spin-$\tfrac{1}{2}$ edge states~\cite{mishra2021,hagiwara1990}. Since the TSCs are grown on a metallic [Au$(111)$] surface, the spin-$\tfrac{1}{2}$ edge states undergo Kondo screening~\cite{Hewson_1993} observed as Kondo resonances near the triangulene terminal units~\cite{mishra2021}. In addition, recent experiments indicate that thin, pristine Ag$(111)$ and Au$(111)$ films can be proximitized by bulk superconductors~\cite{trivini2023,schneider2024,schneider2023,vaxevani2022,liu2023}, which hints at the possibility of growing TSCs via on-surface synthesis on superconductors.

\textbf{\begin{figure}[b]
    \centering
    \includegraphics[width=0.5\textwidth]{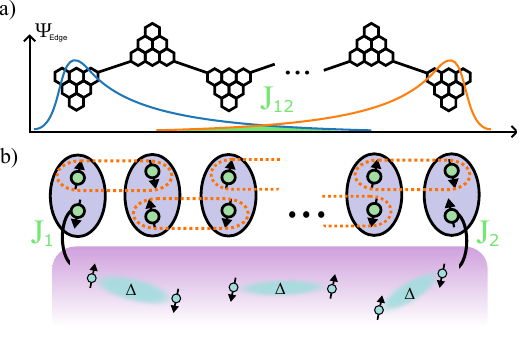}
    \caption{(a) Sketch of an open triangulene spin chain (TSC, see e.g.~Ref.~\onlinecite{mishra2021}). For sufficiently long chains, it hosts two spin-$\tfrac{1}{2}$ states that are exponentially localized near the edges. (b)~Illustration of the TSC on a superconductor. The edge states couple to the superconductor via Kondo exchange interaction. Since the latter is weaker than the antiferromagnetic exchange between the inner spin-$1$ triangulenes,  the central region  remains decoupled in a valence bond solid (VBS) phase. }
    \label{fig:Fig1}
\end{figure}
}
In parallel with these developments, significant progress has been made in spin qubits (SQs) since the seminal work by Loss and DiVincenzo \cite{PhysRevA.57.120}. Various SQ platforms have been explored and continuous improvements in quantum coherence and addressability have been demonstrated~\cite{Kloeffel2013,PhysRevLett.73.2252,Nadj-Perge2010,PhysRevLett.110.066806,Scheibner2008,Gaita-Arino2019}. Despite these advances, decoherence is a major challenge whose main cause is the Overhauser-field noise arising from spin-orbit interaction ~\cite{PhysRevLett.87.207901,PhysRevLett.88.047903} and hyperfine coupling with the nuclei in the quantum dot~\cite{PhysRevB.70.195340,PhysRevB.78.155329}. One possible way out of this conundrum is to design qubits based on spin-singlet states~\cite{Danon2021} which, to leading order, are immune to the Overhauser-field noise.

In this manuscript, we propose a spin-singlet qubit based on a TSC grown on a superconductor; see Fig.~\ref{fig:Fig1}(b).  Unlike a previous proposal based on `Shiba molecules' of magnetic adatoms~\cite{yao2014}, we do not necessarily rely on the Ruderman-Kittel-Kasuya-Yosida (RKKY) interaction, whose sign and strength are difficult to control since it depends on the inter-impurity distance and orientation relative to the substrate. Instead, in the VBS phase of the TSC, the sign (ferromagnetic (FM) or antiferromagnetic (AFM)) and strength of the effective Heisenberg exchange between the edge states are tunable by controlling the chain length~\cite{Affleck1990,mishra2021}. Furthermore, the strength of the super-exchange interaction estimated both experimentally~\cite{mishra2021} and theoretically~\cite{Saleem_et_al_NanoLett_2024} is much larger than the typical RKKY strength~\footnote{ $J_\text{RKKY}\sim  \Delta e^{-2L/\xi_\text{SC}} \frac{\cos^2(k_FL)}{2(k_FL)^2}$~\cite{PhysRevLett.113.087202RKKY}. Taking some realistic values, corresponding to a Nb sample, such as $\Delta\sim2.5$~meV, $\xi_{SC}\simeq 40$~nm, and $k_F\sim13.7 \text{nm}^{-1} $, we have $J_\text{eff}/J_\text{RKKY}\gg1$ for the experimentally observed triangulene chains (5 to 15 units)}.

Moreover, since the exchange between the inner triangulenes ($\sim 10$~meV~\cite{mishra2021,Saleem_et_al_NanoLett_2024}) is stronger than their effective Kondo exchange with the substrate, the coupling of the inner units to the substrate can be neglected,  as experimentally demonstrated in Ref.~\onlinecite{mishra2021}. Thus, we can model the TSC as a two-Kondo-impurity system with the impurities representing the spin-$\tfrac{1}{2}$ edge states. In App.~\ref{app:beyondtwoimp}, by studying a full spin-$1$ chain coupled to a model superconductor using the density-matrix renormalization-group (DMRG),  we provide numerical evidence that supports this two-impurity approximation. Furthermore, due to the van der Waals interaction between the molecule and an STM tip, it is possible to tune the Kondo exchange of the terminal triangulene units~\cite{franke2011,trivini2023,karan2024,Niels2024}, and thus drive the system into a regime in which an avoided crossing of two lowest-lying spin-singlet states occurs. Alternatively, by studying shorter (longer) TSCs, where the Heisenberg (Kondo) exchange dominates, it is possible to observe the system on either side of the avoided level-crossing, provided the edge states are sufficiently strongly coupled to the substrate~\footnote{This regime may be accessible in open ring TSCs for which the edge distance is small enough for the Kondo exchange of both terminal units to be tuned using a single STM tip.}. 

\begin{figure}[t]
    \centering
    \includegraphics[width=0.98\linewidth]{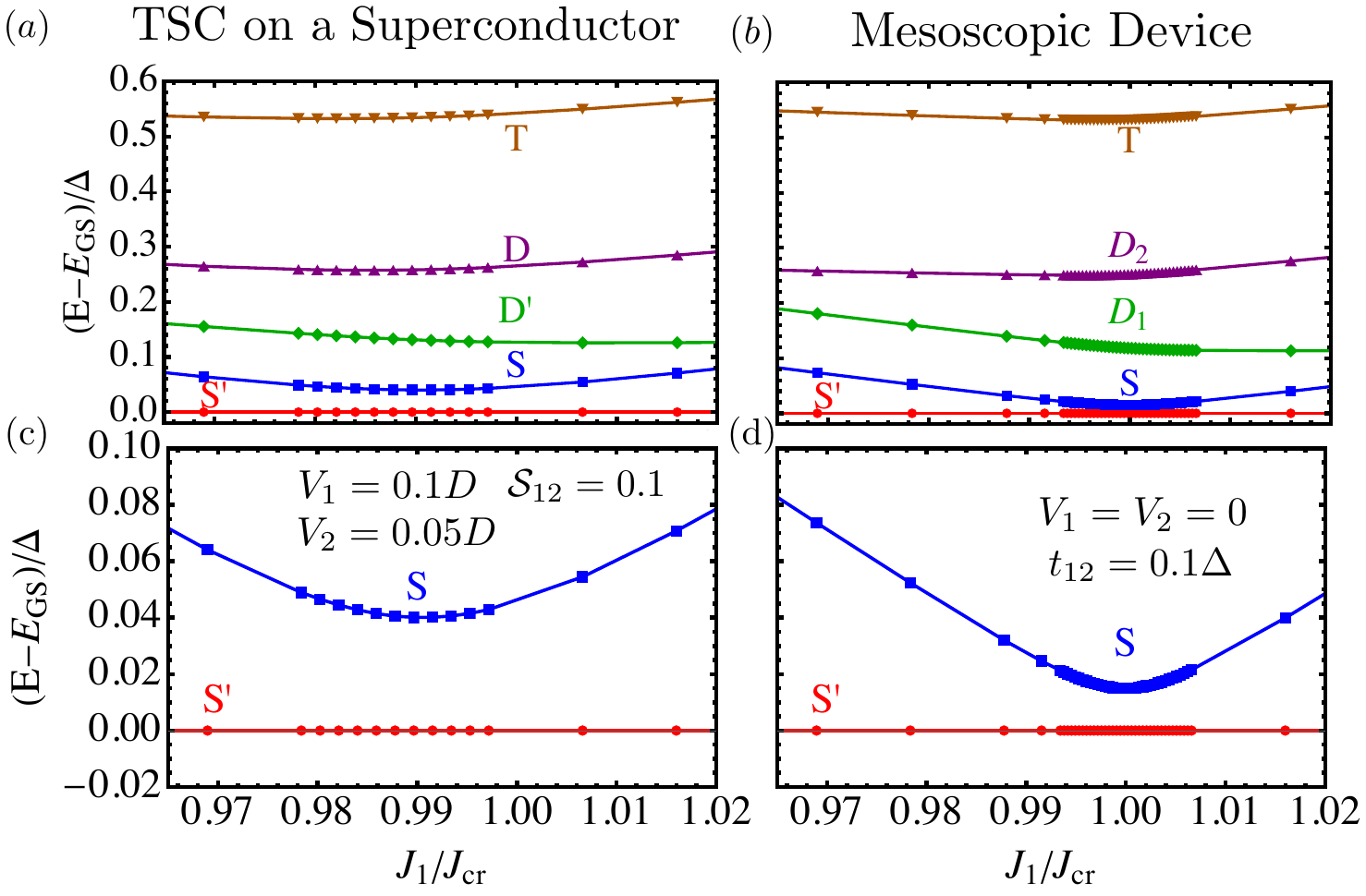}
    \caption{Low-lying spectrum for $J_{12}=\Delta=0.025D$ and $J_2=20\Delta$: (a)  Low-energy spectrum for the effective two Kondo-impurity model describing a triangulene spin chain (TSC) on a superconductor (cf. Fig.~\ref{fig:Fig2}). (b) Low-energy spectrum of the mesoscopic device that ``quantum simulates'' the TSC on a Superconductor (cf. Fig.~\ref{fig:fig3}) (c) Blow-up of the avoided crossing of spin-singlet states for the TSC. The minimum gap between the spin singlets is controlled by the scattering potentials $V_1$ and $V_2$ (d) Blow up of the avoided crossing for the device, in which minimum gap can be controlled by the junction tunneling amplitude  or phase bias.}
    \label{fig:Fig2}
\end{figure}

As pointed out above, using spin singlets suppresses decoherence due to Overhauser fields~\cite{PhysRevLett.87.207901,PhysRevLett.88.047903,PhysRevB.70.195340,PhysRevB.78.155329,Danon2021}. The superconducting nature of the platform provides further protection through the energy gap, enhancing the overall robustness and quantum coherence of the qubit. However, using current STM technology to probe and tune the TSCs may not allow for a practical operation of the spin-singlet qubit. For this reason, we
propose below a mesoscopic
realization using a triple quantum dot  (see Fig.~\ref{fig:fig3}), which exhibits the same low-lying spectrum as the TSC-superconductor system  (Fig.~\ref{fig:Fig2}(a)) and in this sense,  the device is used to ``quantum simulate'' the TSC system,
with the added advantage that the operation and readout of the qubit are feasible using current state-of-the-art electronics~\cite{Sanchez2014,baart2017}.

The rest of this article is organized as follows: In the following section (Sec.~\ref{sec:model}), we introduce a unified two-impurity model that describes both the TSC molecules and the triple
quantum-dot device coupled to superconductors.  Sec.~\ref{sec:avoided} focuses on the avoided crossing of the two lowest-lying (spin-singlet) states  and its description with a simplified 
zero-bandwidth model. In Sec.~\ref{sec:device}, we introduce the quantum-dot device that ``quantum simulates'' the spin-singlet qubit in the TSC molecule-superconductor system and allows for an easier operation of the latter.  In Sec.~\ref{sec:dyn} an effective two-level model of the spin-singlet qubit is introduced. The  time dynamics of the device as obtained using time-dependent NRG are compared to calculations using the two-level model. Our results demonstrate that the qubit can be driven and read by applying voltage pulses to the various gates of the device. The conclusions of this work are given in Sec.~\ref{sec:concl}. Finally, the appendices contain  more in-depth discussions on applicability of the zero-bandwidth approximation, the approximations required to arrive at the  unified model of Sec.~\ref{sec:model}, and the relaxation and  decoherence effects on the qubit device, as well as technical details of our implementation of the NRG.

\section{Model}\label{sec:model}
\begin{figure}[b]
    \centering
    \includegraphics[width=\columnwidth]{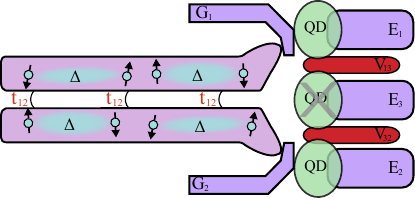}
    \caption{Sketch of the mesoscopic device that emulates the TSC-superconductor system using a triple quantum dot system with the outer dots coupled to a junction of two superconducting wires. Seven gates, $G_1$, $G_2$, $E_1$, $E_2$, $E_3$,  $V_{13}$ and $V_{32}$ define the dots and control the system parameters, including the Kondo couplings, $J_1$ and $J_2$, and the (AFM) super-exchange coupling $J_{12}$. The tunneling amplitude $t_{12}$ or the phase bias across the junction determines the size of the minimum gap at the avoided crossing of the two lowest-energy spin-singlets.}\label{fig:fig3} 
\end{figure}

Within the two-impurity approximation described above, the Hamiltonian of the TSC-superconductor (Fig.~\ref{fig:Fig1})  and the mesoscopic device (Fig.~\ref{fig:fig3})  can be  written down in a unified way as follows:
\begin{align}\label{eq:HeS}
H&= H_0 + \sum_{\nu}J_\nu\bold{S}_{\nu}\cdot \bold{s}_\nu +J_{12} \bold{S}_1\cdot\bold{S}_2\notag\\
&+V_\nu \sum_\sigma\int d\epsilon \: \rho(\epsilon) a^\dag_{\epsilon\sigma\nu}a_{\epsilon\sigma\nu},\notag\\
H_0&=\sum_{\nu,\nu^{\prime}} \int d \epsilon  \: \rho(\epsilon) \biggl[ \sum_{\sigma} \: \epsilon \mathcal{S}_{\nu \nu^{\prime}}(\epsilon) \: a^\dag_{\epsilon\sigma\nu} a_{\epsilon\sigma\nu^\prime}\notag \\&+\Delta\delta_{\nu\nu^{\prime}}\left( a^\dag_{\epsilon\uparrow\nu} a^\dag_{\epsilon\downarrow\nu^{\prime}} + \text{H.c.}\right)\biggr],
\end{align}
where $J_\nu$ and $V_\nu$ are, respectively, the Kondo couplings and scattering potentials of the impurity at ${\bold r} = {\bold r}_{\nu}$ ($\nu=1,2$); $J_{12}$ is the (Heisenberg)  exchange between the two impurities (edge states); ${\bold S}_{\nu}$ is the impurity spin operator and ${\bold s}_{\nu}$ the spin of the itinerant electrons at ${\bold r} =  {\bold r}_{\nu}$. In addition,  $\rho(\epsilon) = \rho_0$ is the density of states in the normal state, which is assumed to be a constant $\rho_0 = 1/(2D)$ ($D$ is the bandwidth).  For a TSC on a superconductor, the overlap matrix $S_{\nu\nu^{\prime}}(\epsilon)$ describes the quantum  amplitude for an electron to propagate from   ${\bold r}_{\nu}$ to   ${\bold r}_{\nu^{\prime}}$; hence $S_{\nu\nu}(\epsilon) = 1$. The detailed form of $S_{12}(\epsilon)$ is not important. For instance, it can be obtained assuming that the single-particle states of the host are plane waves~\cite{yao2014}.

\section{Avoided crossing of spin-singlets}\label{sec:avoided}

Panel (a) in Fig.~\ref{fig:Fig2} shows the low-lying spectrum obtained using NRG for an AFM Heisenberg coupling $J_{12} = \Delta$ as a function of one of the Kondo couplings, $J_1$. We note that the two lowest-lying (spin-singlet) states, $S$ and $S^{\prime}$, undergo an avoided level crossing for $J_{1}\simeq J_{\mathrm{cr}}$ ($J_{\mathrm{cr}}/\Delta \simeq 21.26$) (see also panel (b)). This is a consequence of the breaking of particle-hole symmetry, as anticipated in Ref.~\onlinecite{yao2014}. 
That work studied an RKKY-coupled `Shiba molecule' with  $J_1 = J_2$ and did not report explicit numerical results for the avoided crossing in the case $J_1\neq J_2$  and $V_{\nu=1,2}\neq 0$. Besides the spin singlets, spin doublets, $D$ and $D^{\prime}$ of opposite fermion parity, and a spin triplet, $T$, are found at higher excitation energies but still below the superconducting gap $\Delta$. Therefore, transitions between the singlets and doublets should be observable as Yu-Shiba-Rusinov~\cite{yu1965,shiba1968,rusinov1969} peaks in the  scanning tunneling spectra at subgap energies. 

Interestingly,  the low-lying spectrum can be reproduced using a zero-bandwidth approximation~\cite{vonoppen1} (ZBA, see  App.~\ref{appA} for additional details), which \textit{effectively} captures the breaking of the particle-hole symmetry. In the ZBA, the superconductor Hamiltonian ($H_0$ in Eq.~\eqref{eq:HeS}) is replaced by two fermion sites with a pairing potential: $\Delta \sum_{\nu}c^{\dag}_{\nu \uparrow} c_{\nu \downarrow} + \mathrm{H.c.}$ The  overlap    $\propto S_{\mu\nu}$ of the channel orbitals is described by a hopping term, $H_{12} = -t_{12}\sum_{\sigma} c^{\dag}_{1\sigma} c_{2\sigma} + \mathrm{H.c.}$~\cite{schmid2022}. For $t_{12}=0$,  the lowest-lying states are two (valence-bond) spin-singlets $S_0$ and $S_2$ (see table~\ref{tbl:states}) that cross for $J_1  = J_{1cr}$ (see Fig.~1(b) and 1(d) in SI). However,  for $t_{12} > 0$,  an avoided crossing occurs between the two spin-singlet states, $S$ and $S^{\prime}$, which are linear combinations of $S_0$ and $S_2$ (see  App.~\ref{appA} for extended discussion of the ZBA).

Summarizing the results in this section, in the parameter regime shown in Figs.~\ref{fig:Fig2}(a) and (c) an isolated manifold of two lowest-lying states undergoing an avoided  realizes a spin-singlet qubit in the TSC-superconductor system.  The qubit could be mechanically operated by tuning the Kondo coupling at one end of the TSC using the STM~\cite{franke2011,trivini2023,karan2024,Niels2024}. However, the  frequencies at which the tip position can be modulated with current STM technology are several orders of magnitude smaller than the minimum gap between $S^{\prime}$ and $S$ ($\approx \Delta/100 \approx 0.01$ meV $\approx 1$ GHz). Thus,  creating quantum superpositions of $S$ and $S^{\prime}$ by mechanical driving may be rather challenging, not to mention performing the qubit readout~\footnote{An antenna emitting a microwave field could be used to control the qubit, similar to electron-spin resonance STM or atomic-force microscope (AFM) measurements in Refs.~\cite{Baumann2015,Sellies2023}. However,  new detection mechanisms would be required, as the singlet nature of the qubit prevents the readout via magnetoresistance changes~\cite{Baumann2015}. Charge state variations could still be used~\cite{Sellies2023}}. Instead,  we propose below a mesoscopic device with the same low-lying spectrum in which the system parameters can be electrically controlled and driving and readout are possible using state-of-the-art electronics.

\begin{table}[t]
  \centering
  \begin{tabular}{  l | l | c  }
   \hline
   state& ($S$, $P$) & \qquad\quad figure \\ \hline
   Impurity singlet ($S_0$) &    $(0, +1)$   &
    \qquad\quad\begin{minipage}{.1\textwidth}
     \centering
      \includegraphics[height=0.5cm]{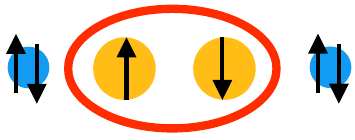}
    \end{minipage}
    \\ 
     Kondo singlet ($S_2$) &    $(0,+1)$   &
  \qquad\quad \begin{minipage}{.1\textwidth}
     \centering
      \includegraphics[height=0.5cm]{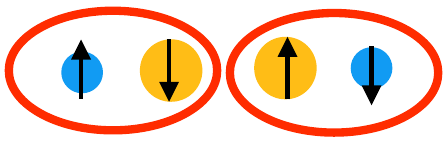}
    \end{minipage}
    \\ 
 Doublets ($D_1,D_2$) &    $(\frac{1}{2}, -1)$   &
   \begin{minipage}{.1\textwidth}
   \centering
      \includegraphics[height=0.5cm]{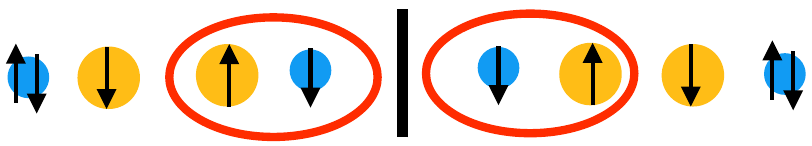}
    \end{minipage}
    \\ 
Triplet ($T$) &    $$(1, +1)$$   &
   \qquad\quad \begin{minipage}{.1\textwidth}
      \centering
     \includegraphics[height=0.5cm]{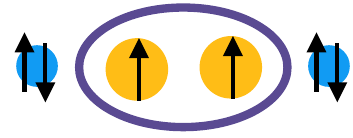}
   \end{minipage}
  \end{tabular}
  \caption{Lowest-energy states in the zero-bandwidth approximation; $S$ and $P = (-1)^{N_F}$
  are the spin and fermion parity quantum numbers, respectively ($N_F$ is the total fermion number in the superconductor). Red and blue solid circles represent spin-singlet and spin-triplet states, respectively. Blue (orange) dots represent the Kondo-impurity (superconductor) sites.
    }\label{tbl:states}
\end{table}

\section{Mesoscopic device}\label{sec:device}

Since the ZBA is capable of reproducing the low-lying spectrum of the TSC-superconductor system using a tunneling term (compare Fig.~\ref{fig:s1} in App.~\ref{appA}  to Fig.~\ref{fig:Fig2}(a)), we use superconducting wires forming a tunneling junction. Moreover, the edge states of the TSC can be replaced by a triple quantum dot~\cite{Sanchez2014,baart2017}. The resulting device (see Fig.~\ref{fig:fig3}) is described by Eq.~\eqref{eq:HeS} with $\mathcal{S}_{12} = -t_{12}/\epsilon$ and $\mathcal{S}_{\nu\nu} =1$. The triple quantum dot is a minimal setup to engineer the AFM coupling via super-exchange $\propto J_{12} > 0$ in Eq.~\eqref{eq:HeS}. In such systems, $J_{12}$  ranges from  $0.01$ to $0.1$ meV~(see e.g. Refs.~\onlinecite{Deng_PhysRevB.102.035427,baart2017,Sanchez2014,Sanchez2014b}).  Roughly speaking, the charging energy of the central dot suppresses single-particle tunneling between outer quantum dots. This results in the low-lying level structure shown in Figs.~\ref{fig:Fig2}(b,d), where the two spin-singlets are the lowest-lying states, well separated from the spin-doublet and triplet states. NRG finds an avoided crossing between  the spin singlets. The minimum gap can be controlled by $t_{12}$ or the phase bias between the superconductors rather than $V_{1,2}$, as it is the case of the TSC-based system. In App.~\ref{AppB}, we show that accounting for charge fluctuations within the Anderson model does not spoil these spectral features.

\begin{figure*}[t]
    \centering
    \includegraphics[width=\textwidth]{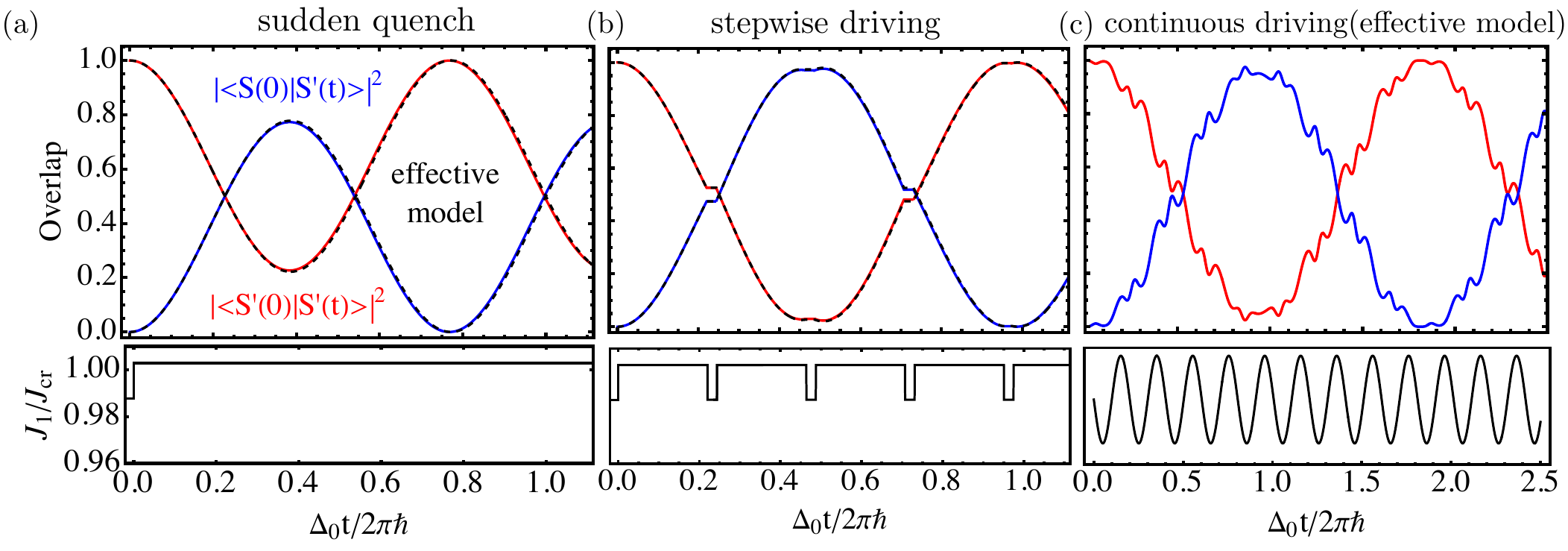}
    \caption{Panels (a) and (b): Blue and Red lines are the overlaps between the time-evolved qubit state and the two initial singlet states computed using time-dependent NRG. The black dashed lines correspond to the overlap derived from the effective two-level model in Eq.~\eqref{eq:tls}. The figures below are the corresponding quench profiles. (c) Overlap under a continuous driving calculated from the two-level effective model from Eq.~\eqref{eq:tls}.}\label{fig:sudden}
\end{figure*}

It is worth emphasizing that the  spin-doublet states of opposite fermion parity  (cf. table~\ref{tbl:states}) lie above the singlets, and this energy separation makes our qubit proposal somewhat more robust against quasi-particle poisoning than earlier proposals where the doublets lie in between the singlets~\cite{steffensen2024ysrbondqubitdouble_bond_q}~\footnote{Our NRG calculations of double quantum-dot system coupled by tunneling (rather than super-exchange) confirm the results of Ref.~\onlinecite{steffensen2024ysrbondqubitdouble_bond_q}, which were obtained in the infinite gap limit and yield a low-lying spectrum where the odd fermion-parity doublets lie between the even fermion-parity singlets~\cite{unpub2025}.}. Gap engineering~\cite{pan_engineering_2022} and the implementation of quasi-particle traps~\cite{PhysRevB.94.104516} can be further deployed to enhance the qubit's resilience against quasiparticle poisoning. However, a reliable quantitative analysis of these effects requires detailed knowledge of the specific circuit in which the qubit is embedded~\cite{PhysRevB.89.104504,PhysRevB.72.134519}. Therefore, precise estimates of quasi-particle poisoning lie beyond the scope of the present work.


\section{Dynamics: Driving and Readout}\label{sec:dyn}

Next, we develop and test an effective two-level model that describes the qubit dynamics as realized in the mesoscopic device introduced above (Fig.~\ref{fig:fig3}). The dynamics discussed below are assumed to be unitary, and therefore neglect relaxation and dephasing  (see App.~\ref{app:dephasing} for a discussion) and  quasi-particle poisoning, upon which we have already touched above. In the parameter regime shown in Figs.~\ref{fig:Fig2}(c,d) with fixed $J_2,\Delta,V_1,V_2$ and $J_{12}$, the following effective two-level Hamiltonian can be written down:
\begin{align}
  H_{\text{eff}} &=  \left[\epsilon_0 +  \alpha (J_1(t)-J_0)\right] |S_2\rangle \langle S_2 |\notag\\
 &+ \frac{\Delta_0}{2}\left( |S_0\rangle \langle S_2 |+ |S_2\rangle \langle S_0 | \right). \label{eq:tls}
\end{align}
The basis of spin-singlet states $\{|S_0\rangle, |S_2\rangle\}$ is defined for zero tunneling amplitude at the junction, i.e. $t_{12} = 0$, for which there is no avoided crossing. The parameter $J_1(t)$ is the exchange coupling tuned by the gate $G_1$ (see Fig.~\ref{fig:fig3}) that controls the tunneling ($t_1$) between the upper quantum dot and  the closest superconducting wire of the junction ($J_1 \propto t^2_1$, see e.g. Ref.~\onlinecite{Hewson_1993}); $\epsilon_0=E_{S_2}-E_{S_0}$ is the energy difference between the singlets for $t_{12}=0$; $J_1(t=0)=J_0$ is the Kondo exchange of a reference (initial) state. We assume a linear dependence of $\epsilon_0$ on $J_1$ for $J_1\simeq J_0$ and define $\alpha=\frac{\partial \epsilon_0(J_1) }{\partial J_1} |_{J_1=J_{0}}$ in terms of the energy of $|S_2\rangle$ at $t_{12}=0$ to account for it. The off-diagonal term $\Delta_0$ describes the coupling between the spin-singlet states induced by a finite $t_{12}$. All the parameters are obtained using NRG.

We have tested the effective two-level model in Eq.~\eqref{eq:tls}  by comparing it to time-dependent NRG simulations of the full system (cf. Eq.~\eqref{eq:HeS}) in various scenarios. First, we consider a quantum quench where $J_1(t)$ suddenly changes. In Fig.~\ref{fig:sudden}(a) we assume the system is prepared in the ground state $|S'\rangle$ with $J_1/J_{cr} = 0.988$ and the Kondo coupling $J_1$ is suddenly quenched to $J_1/J_{cr}=1.003$. We thus compute the overlap of the time-evolved state with the two initial spin-singlet states both using the effective model \eqref{eq:tls} and time-dependent NRG for the full model. The effective model agrees well with the time-dependent NRG results. It is worth noticing that a sudden quench of $J_1(t)$ conserves the total spin and, therefore, only states with the same spin as the initial state (i.e., spin-singlets) can be excited during the time evolution. As the lowest-lying spin-singlets are well separated from other continuum states with zero total spin, the quench causes minimal admixture and the dynamics are well reproduced by the two-level model,  Eq.~\eqref{eq:tls}.  We have also considered more complex types of quenches. In Fig.~\ref{fig:sudden}(b),
a stepwise drive where $J_1(t)/J_{cr}$ oscillates between $0.988$ and $1.003$ with period $T=40.98\times2\pi\hbar/\Delta$. The results from Eq.~\eqref{eq:tls} accurately track the time-dependent NRG results.  

In the following, we use the two-level model, Eq.~\eqref{eq:tls}, to show that Rabi oscillations between qubit states can be induced by a simple periodic modulation of one of the Kondo couplings, with $J_1(t)  = J_1(0)+ \delta J\sin(\omega t)$ for a fixed $J_2$. The corresponding oscillations are shown in Fig.~\ref{fig:sudden}(c), where $J_1(t)/J_{cr}$ oscillates between $0.988 \pm 0.0188$ with period $T=2\pi\hbar/\epsilon_0= 33.87 \times 2\pi\hbar/\Delta$. To avoid admixing  high-energy states, the driving frequency must be much smaller than the superconducting gap: $\omega= \epsilon_0\ll \Delta/2\pi\hbar$.  The initial state is the ground state for $J_1/\Delta=21$ $(J_1/J_{cr}=0.988)$. This driving protocol can be implemented in the setup shown in Fig.~\ref{fig:fig3} using the gate electrodes connected to the quantum dots ($V_{1}$ and $V_{2}$ in Fig.~\ref{fig:fig3}), which govern the tunneling into/from the outer quantum dots, effectively modulating the Kondo exchange $J_{1,2}$. An oscillating electric field applied to the gate electrode could serve as the driving force. Notably, in this setup, $J_1$ and $J_2$ can be independently controlled, though this is not strictly necessary for qubit operation, as modulation of both couplings simultaneously is also viable.
Readout can be achieved by measuring the absorption spectrum of an applied AC field~\cite{janvier2015,hays2021}. Absorption is maximized at the transition frequency, enabling precise characterization of the qubit state. Similar to Ref.~\onlinecite{steffensen2024ysrbondqubitdouble_bond_q}, the device in Fig.~\ref{fig:fig3} can  be integrated into a cavity, allowing the readout of the qubit by coupling to the cavity modes.

\section{Conclusions}\label{sec:concl}

In conclusion, we have theoretically proposed a novel qubit platform that can be realized using a TSC grown on a superconducting substrate. However, to realistically operate and read out the qubit, we have also introduced a mesoscopic device that quantum simulates the TSC-superconductor system. We have identified a parameter regime where the qubit has two lowest-lying spin-singlet states, which make the qubit immune to random magnetic fields. Quasi-particle poisoning~\cite{10.1063/PT.3.5291,PhysRevLett.72.2458,PhysRevLett.92.066802} can still be a source of decoherence, giving rise to a non-zero population of the doublets $D_1$ and $D_2$. However, the clear separation in energy of the singlet states and the doublets is expected to provide additional protection against the latter. Known schemes to avoid poisoning  can be  also implemented~\cite{PhysRevLett.103.097002,PhysRevLett.92.066802,PhysRevB.94.104516,pan_engineering_2022}.  Finally, it is also worth mentioning that the simple readout mechanism outlined in the previous paragraph has been successfully used in state-of-the-art transmon-like qubits~\cite{janvier2015,hays2021}.

\begin{acknowledgments}

We acknowledge useful discussions with F. S. Bergeret  J. I. Pascual, S. Trivini, R. Aguado, and A. Levy-Yeyati. This work has been supported by the Agencia Estatal de Investigaci\'on (AEI) of the Ministerio de Ciencia, Innovaci\'on, y Universidades (Spain) through AEI/10.13039/501100011033 Grants No. PID2020- 120614GB-I00 (ENACT) and No. PID2023-148225NB-C32 (SUNRISE). J. O. acknowledges the scholarship PRE\_2021\_1\_0350 from the
Basque Government.

\end{acknowledgments}

\section*{DATA AVAILABLITY}
The data that support the findings of this article are openly available~\cite{zenodo}.

\appendix
\section{Zero-Bandwidth Approximation (ZBA)}\label{appA}

\begin{figure*}[t]
    \centering
    \includegraphics[width=0.95\linewidth]{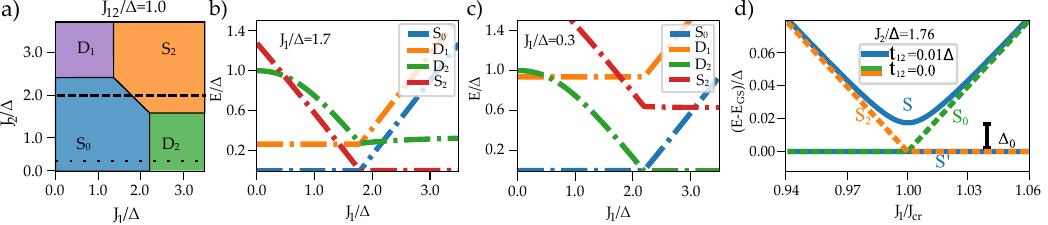}
    \caption{\textbf{Results obtained using the zero-bandwidth approximation (ZBA):} (a) Phase diagram of the two impurity model for $J_{12}=\Delta$ and $t_{12}=0$ as a function of the individual Kondo exchange couplings with the substrate, $J_1$ and $J_2$. (b) and (c) Evolution of the energy of the four lowest-lying states along the dashed and dotted lines in panel (a), respectively. (d) Comparison between no hopping (dashed lines) and a small hopping term (solid line) in the vicinity of the $S_0\rightarrow S_2$ transition in panel (a).}
    \label{fig:s1}
\end{figure*}

The ZBA \cite{vonoppen1,ortuzar2023,schmid2022} can be used to qualitatively describe the low-lying spectrum of the systems studied in the main text, as obtained using NRG. In the ZBA, the Hamiltonian reads:
\begin{align}\label{Hs}    
&H=H_{0}+H_S+H_t\\
&H_0= \Delta \sum_{\alpha=1,2}\left[  c^{\dagger}_{\alpha\uparrow}c^{\dagger}_{\alpha\downarrow}+ \mathrm{H.c.}\right]\\
&H_S=J_{12}\bold{S}_1
\cdot \bold{S}_2+\sum_{\alpha=1,2}J_\alpha  \bold{S}_\alpha \cdot \left( c^{\dagger}_{\alpha s} \boldsymbol{\sigma}c_{\alpha s^{\prime}} \right) \\
&H_t=- t_{12} \sum_{s} \left[e^{i\phi/2} c_{1\sigma}^\dag c_{2\sigma} + \mathrm{H.c.}\right],
\end{align}
where, $c^{\dagger}_{\alpha s}$ ($c_{\alpha s}$) is the creation (annihilation) operator of an electron with spin $s$ on the $\alpha$ site, $\Delta$ is the superconducting pairing energy, $J_{\alpha}$ describes the Kondo exchange coupling of the edge states with the substrate, $J_{12}$ is the exchange interaction between the edge state spins, and $t_{12}$ is a hopping term between the superconductors, which accounts for the extension of the YSR states~\cite{schmid2022} and the overlap between the channels that couples to the edge states. A phase bias, $\phi$, controlled by an external flux, has been added to the hopping term. It determines the size of the energy difference between the lowest-lying spin-singlet states. Thus, along with the hopping amplitude of the junction (described by $t_{12}$ in the above model), the phase bias $\phi$ can also be used to control the gap at the avoided crossing. However, noise fluctuations in $\phi$ can  be a  source of dephasing (see Fig.~\ref{re_de} below and discussion in App.~\ref{app:dephasing}).

The lowest-lying states described by the ZBA Hamiltonian for $t_{12} = 0$ are listed in Table~$1$ in the main text. The phase diagram, as a function of the Kondo exchange  between the edge states and the substrate, is shown in Fig.~\ref{fig:s1}(a). Notably, for finite $J_{12}$, a transition between the two lowest-lying spin-singlet states $S_0$ and $S_2$ takes place along the trajectory in parameter space indicated by the black dashed line. The evolution of the energy of the four lowest-energy states along this trajectory is shown in Fig.~\ref{fig:s1}(b), where a (level-crossing) quantum phase transition occurs between  $S_0$ and $S_2$, while the doublet states remain well separated at higher energy. This transition can be regarded as a non-local quantum phase transition, since tuning one of the Kondo exchange couplings of one impurities drives the entire system across a phase transition  into a different (spin-singlet) ground state in which both impurities capture a quasi-particle from their respective superconducting baths (recall that $t_{12} = 0$ here and the two superconducting sites are not coupled and therefore independent).

Additionally, there is a  parity-changing quantum phase transition~\cite{yu1965,shiba1968, rusinov1969,vonoppen1}. The latter is accessible along the dotted black line shown in Fig.~\ref{fig:s1}(a). The corresponding  spectrum evolution is shown in Fig.~\ref{fig:s1}(c), highlighting the parity change of the ground state between the (spin-singlet) $S_0$ state and the (spin-doublet) $D_2$ state.  

As briefly mentioned in the main text, the hopping term $H_{12}$ \emph{effectively} breaks global particle-hole symmetry, mixing the two singlet states by generating a finite matrix element between $S_0$ and $S_2$.
Global particle-hole symmetry (PHS) is defined by the transformation where $c_{\nu\sigma} \to c^{\dag}_{\nu,-\sigma}$, which changes $H_{12}\to -H_{12}$. Thus, setting $t_{12} \neq 0$ breaks the global PHS, and results in the avoided crossing of the two singlets. Let us recall that, within the ZBA,  $H_{12}$ is intended to describe the overlap between the  channels that couple to the two different edge states of the TSC chain, namely the term $\propto S_{12}(\epsilon)$ in Eq.~\eqref{eq:HeS} of the main manuscript.  However, it can be shown (see App.~\ref{AppC} below) that the choice $S_{12}(\epsilon) = 1$  describing the two-impurity system on the \emph{same} superconductor, \emph{does not} break  global particle-hole symmetry. Indeed, the latter is  broken by 
the scattering potentials $\propto V_{1}, V_{2}$ in Eq.~\eqref{eq:HeS}.  Therefore,  in this sense, the  ZBA  must be regarded as an \emph{effective} description and we must carefully compare it with the results obtained for the full two-impurity model studied using NRG.

With the above caveat in mind, we notice nonetheless that $H_{12}$ mixes
the spin-singlet states $S_0$ and $S_2$, which are eigenstates at $t_{12} = 0$. This results in a new pair of spin-singlet states, $S$ and $S^{\prime}$, which exhibit an avoided crossing, as shown in Fig.~\ref{fig:s1}(d) ($\phi=0$ in the figure). This figure resembles the results obtained by NRG calculation (see Fig.~\ref{fig:Fig2}, panels (a) and (c)), demonstrating that the ZBA can capture the behavior of the low-energy  spectrum  obtained from NRG as a function of the Kondo exchange of one of the impurities.

  Finally, we notice that the effect of the phase $\phi$ is illustrated in Fig.~\ref{re_de}(a), which shows that an external magnetic field can be also used to control the transition energy of the qubit states in the quantum-dot device (note that ZBA also applies to this system).

\section{Level crossing between the two sub-gap singlet states}\label{app:NLQPT}
\begin{figure}[t]
    \centering
\includegraphics[width= \columnwidth]{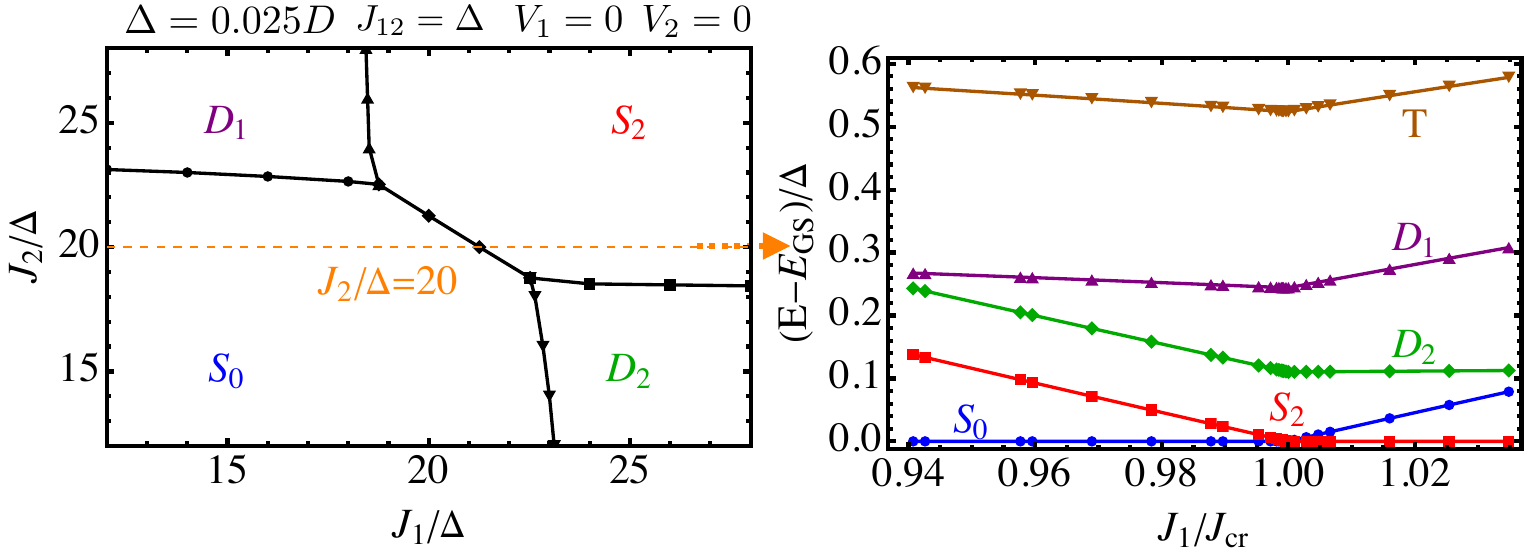}
    \caption{ 
   NRG results for Eq.~\eqref{eq:HeS} of the main text with $S_{12}=0$, i.e. for two independent superconducting baths. Panel (a)  Phase diagram for $J_{12}=\Delta,V_1=V_2=0$. Panel (b) Sub-gap energy levels related to the ground state along the line $J_2/\Delta=20$. }\label{fig:NLQPT}
\end{figure}
 In this Appendix, we briefly discuss the low-energy energy spectrum obtained from NRG when setting  $\mathcal{S}_{12}=0$ and $V_{1,2} =0$ in Eq.~(1) of the main text. The Wilson chain Hamiltonian reads, $H=H_\text{imp}+\tilde{H}_1+\tilde{H_2}$ and it is described in Eq.~\eqref{eq:wc} below. 

\begin{figure}[b]
    \centering
    \includegraphics[width=0.9\linewidth]{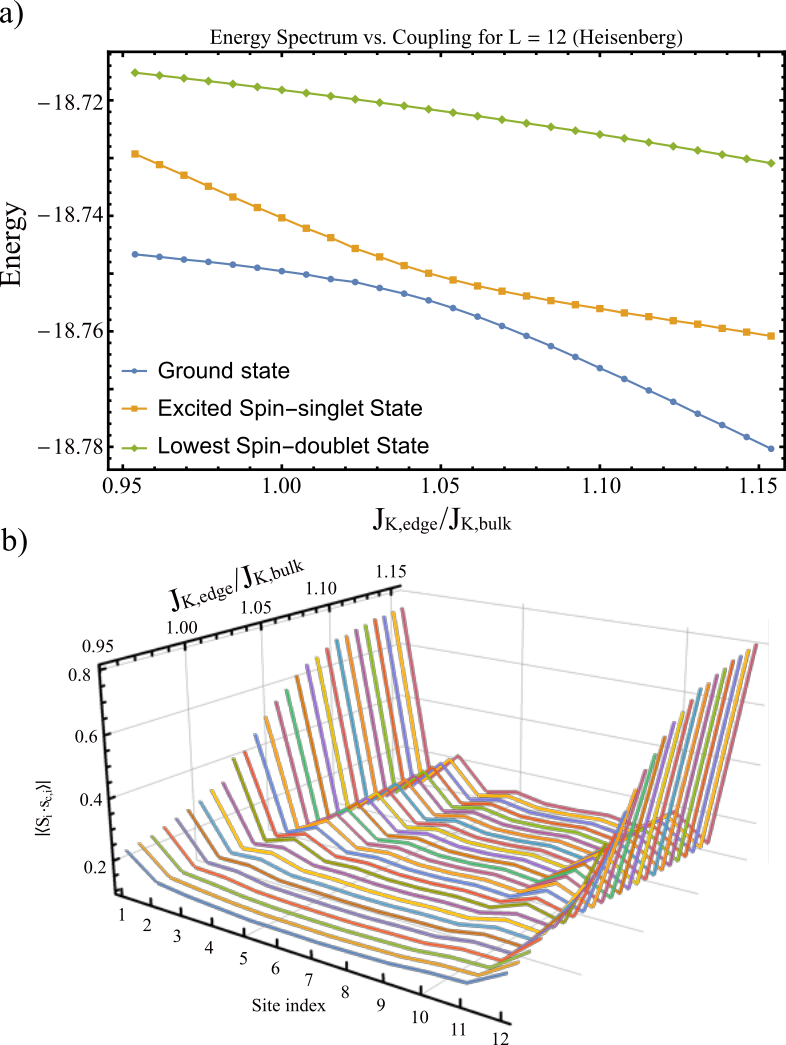}
    \caption{Panel (a): Spectrum of low-lying states as a function of the ratio of the rightmost Kondo coupling  $J_{K,i=12} = J_{K,\mathrm{edge}}$ to the bulk Kondo exchange $J_{K,i=1,\ldots,11} = 1.626\Delta$ for the model in Eq.~\eqref{H_chain} ($\Delta$ is the strength of the pairing potential in the superconducting chain). Panel (b) Absolute value of the impurity impurity-superconductor spin correlations, $\langle \boldsymbol{S}_i \cdot \boldsymbol{s}_{c,i}\rangle$, where $\boldsymbol{s}_{c,i} = \sum_{s,s^{\prime}} c^{\dagger}_{i s} \boldsymbol{\sigma}_{ss'}c_{i s^{\prime}}$ for all sites $i=1,\ldots,12$, as a function of $J_{K,\mathrm{edge}}/J_{K,\mathrm{bulk}}$. Notice the substantial increase in the correlations in \emph{both} edges as the Kondo coupling is increased in the rightmost edge only.}
    \label{SI_chain}
\end{figure}
 
 In this case, as shown in panel (a) of Fig.~\ref{fig:NLQPT}, a phase boundary  between two phases adiabatically connected to the states $S_0$ and $S_2$  shown in Tab.~\ref{tbl:states} exist. These  two spin-singlet states become degenerate at this boundary as shown  in panel (b) of the same figure. The latter displays the evolution of the low-energy spectrum along the path shown in panel (a) as a dashed line, along which $J_2/\Delta=20$ while $J_1$ is varied.   This level crossing is  the nonlocal quantum  phase transition, which was mentioned above in our discussion of the ZBA. 

\section{Beyond the two-impurity approximation}
\label{app:beyondtwoimp}

 In order to assess the role of the intra-chain exchange between the triangulene units in a TSC,
 we have studied an extension of the two-impurity ZBA model using density-matrix renormalization-group (DMRG). This allows us to investigate whether disregarding all internal degrees of freedom of the spin-$1$ chain and replacing it with two spin-$\tfrac{1}{2}$ impurities representing the edge states is a valid approximation for describing the low-lying  spectrum of the TSC-superconductor system. Note that, even with DMRG, studying the spectrum of the $N$-spin chain on an extended superconducting system is a daunting task. Thus, taking notice of the success of the ZBA in describing the low-lying spectrum,  we  replace the full extended superconduct by a finite superconducting chain in the model described below.

 Specifically, we study the spectrum of a $N$-site antiferromagnetic Heisenberg spin-$1$ chain, which is a minimal model for the VBS phase, coupled to a one-dimensional superconducting chain. Each spin-$1$ ``impurity'' (representing a triangulene unit of the TSC) is coupled via Kondo exchange $J_K$ to a single superconducting site of the type considered in the previous section. The spins are coupled to each other with an isotropic (i.e. Heisenberg) exchange $J>0$. Finally,  the superconducting sites are coupled with a constant hopping amplitude, as in Eq.~\eqref{Hs}. The Hamiltonian of the resulting model reads:
\begin{equation}\label{H_chain}
\begin{split}
    \hat H =&\sum_{i=1,ss'}^N J_{K,i} \mathbf{S}_i \cdot \left( c^{\dagger}_{i s} \boldsymbol{\sigma}_{ss'}c_{i s^{\prime}}\right) 
    + \sum_{i=1}^N\Delta \left(c^{\dagger}_{i\uparrow}c^{\dagger}_{i\downarrow}+\text{h.c.}\right) \\
    &+\sum_{\langle ij\rangle}J \mathbf{S}_i\cdot \mathbf{S}_j
    -\sum_{\langle i j\rangle, s} t\left(c^{\dagger}_{i,s}c_{j,s}+\text{h.c.}\right).
\end{split}
\end{equation}
The choice of parameters is as follows: For the Heisenberg coupling, we set $J=5\Delta$. For the Kondo exchange, we set $J_{K, i}=J_{K,\text{bulk}}=1.625 \Delta$ for $i= 1,N-1$, while the right-edge Kondo coupling, $J_{K,i=N}=J_{K,\text{edge}}$. We simulate the tuning of the Kondo coupling by an STM tip on top of the rightmost terminal triangulene unit by changing the ratio  $J_{K,\mathrm{edge}}/J_{K,\mathrm{bulk}}$  between $\simeq 0.95$ and $\simeq 1.15$. Using DMRG we have computed the energy  of the lowest-energy states with fixed total spin-$z$ projection ($S^z$) and fermion parity, $P = (-1)^{N_F}$ (where $N_F$ is the number of fermions in the superconducting chain). In addition, we also compute the absolute value of the  spin-spin correlation $\langle \boldsymbol{S}_{i}\cdot \left( \sum_{s,s^{\prime}} c^{\dagger}_{i s} \boldsymbol{\sigma}_{ss'}c_{i s^{\prime}}\right)\rangle$ in the ground state, which is a measure of the strength of Kondo-like correlations between the spin-$1$ impurities representing the triangulene units and the nearest superconducting sites.

\begin{figure*}[t]
    \centering
    \includegraphics[width=0.95\linewidth]{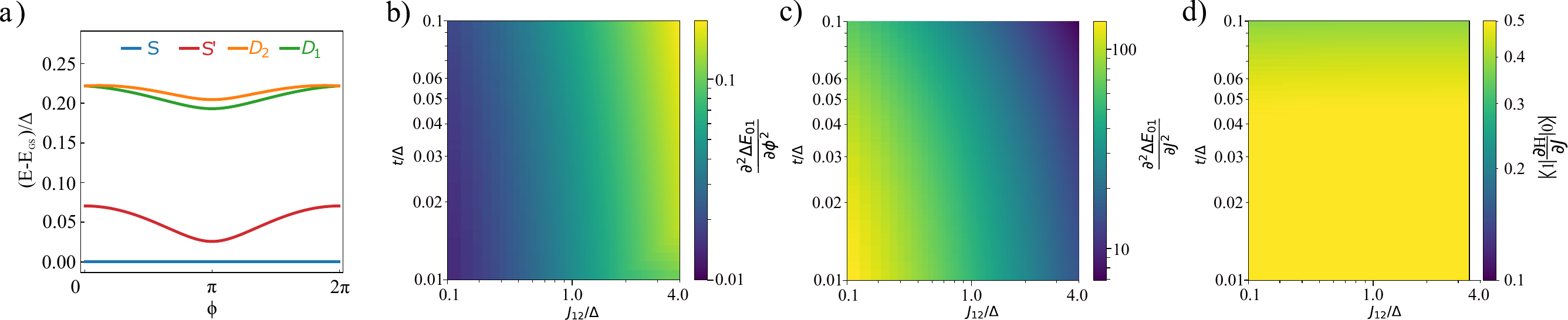}
    \caption{Panel (a)  is the phase  dependence of the energy of the four lowest-lying states of the qubit in the ZBA. Estimations of the optimal working points are shown in panels (b) and (c), which display calculations of the curvature of the qubit states splitting, $E_{01}= E_1-E_0$, with respect to $\phi$ and the Kondo exchange couplings $J_1=J_2$, as a function of the hopping, $t=t_{12}$ and exchange coupling between dots, $J_{12}$. The estimations are calculated for a zero applied external flux. Panel (d) shows the calculated $\left|\bra{1}\frac{\partial \hat H}{\partial J}\ket{0}\right|^2$. In this panel, $J$ corresponds to  one of the two Kondo couplings (i.e. either $J_1$ or $J_2$). }
    \label{re_de}
\end{figure*}

In Fig.~\ref{SI_chain}(a) we show that by tuning the rightmost edge Kondo coupling, the low-energy spectrum displays the same avoided crossing that was observed in the two-impurity model of the chain using ZBA. Both the ground state and the first excited state are spin singlets, while at higher energies we find a spin-doublet of opposite fermion parity, $P$. Examining the spin correlations across the avoided crossing,  Fig.~\ref{SI_chain}(b) shows 
a substantial increase of the latter (by a factor of four) 
as the right-edge Kondo coupling is tuned from $J_{K,\mathrm{edge}} \simeq 0.95 J_{K,\mathrm{bulk}}$ to $J_{K,\mathrm{edge}} \simeq 1.15 J_{K,\mathrm{bulk}}$.
Notice that the spin correlations increase almost symmetrically at both ends of the chain (i.e. for $i=1$ and $i=N=12$). 
However, on average, they remain much smaller at all the other sites (i.e. $i=2,\ldots, 11$) as the system is driven across the avoided crossing of the spin-singlets. 
In terms of the picture provided by the two-impurity ZBA, to the left of the avoided crossing, the ground state resembles more the $S_0$ state (see Tab.~\ref{tbl:states} in the main text) as the spin correlations with the superconductor are weak ($\lesssim 0.2$ typically), despite the fact that $J_{K,\mathrm{bulk}}> \Delta$. However, as we approach the point where the gap between the spin-singlets takes its minimum value (for $J_{K,\mathrm{edge}}/J_{K,\mathrm{edge}} \simeq 1.05$, the correlations between the edge spins  sharply increase up to $\simeq 0.8$, which would be consistent with the formation of a singlet-like states between edge states and the superconductor. In the language of the two-impurity ZBA, this resembles the state $S_2$ (see Tab.~\ref{tbl:states} of the main text).

\section{Relaxation and Dephasing}\label{app:dephasing}

Maintaining quantum coherence is essential for storing and processing information in a qubit. However, interactions with the surrounding environment inevitably lead to decoherence, which occurs through two processes: relaxation and dephasing.

Relaxation, characterized by the qubit lifetime $T_1$, refers to the loss of energy as the system decays from the excited state $\ket{1}$ to the ground state $\ket{0}$ of the qubit. It is typically induced by coupling to environmental modes that can absorb energy, such as photons, phonons, or other excitations. The environmental degrees of freedom appear as noise in a parameter of the system $\eta$, such that the decay process is related to the overlap between the excited- and ground-state wavefunctions $T_1 \propto \left|\bra{1}\frac{\partial \hat H}{\partial \eta}\ket{0}\right|^2$~\cite{PRXQuantum.2.030101,PhysRevB.89.104504,PhysRevB.72.134519,smith_superconducting_2020}. 

Dephasing, characterized by $T_2$, corresponds to the loss of phase coherence between $\ket{0}$ and $\ket{1}$ without necessarily changing their populations. Dephasing arises from random environmental fluctuations (e.g., magnetic, charge, or flux noise) that shift the qubit transition frequency and scramble the relative phase coherence. The dephasing rate depends both on relaxation and on pure dephasing, $1/T_2=1/(2 T_1)+1/T_\varphi$. The latter is related to the curvature of the $0\to 1$ transition energy with respect to external parameters. When a qubit is operated in the optimal working point, the first derivative respect to external parameters vanish, so that the limiting factor for the pure dephasing is given by the second derivative: $\left|\frac{\partial^2 E_{01}}{\partial\eta^2}\right| ^2$~\cite{PRXQuantum.2.030101,PhysRevB.89.104504,PhysRevB.72.134519}.

In this subsection, we estimate the evolutions of the main term affecting to lifetime and dephasing for the triple quantum-dot device discussed in the main text,  
as a function of the external parameters $\phi_{\text{ext}}$ and $J_i$. For this estimation, we use the ZBA. As explained in the main text, this approximation is limited in that we cannot directly relate the model parameters of the calculation with actual experimental ones. However, it yields a simpler model that provides quick access to approximate trends that are useful for identifying the optimal working points of the device as a function of $t=t_{12}$ and $J_{12}$.

Figures~\ref{re_de}(b) and~\ref{re_de}(c) show the second derivative of the qubit transition energy with respect to the external flux and the Kondo coupling, plotted as functions of the hopping $t$ and inter-dot exchange $J_{12}$. The second derivatives are evaluated at the minimum-energy point, where first derivatives vanish, i.e., the optimal working point. We find that the former estimate decrease with decreasing Kondo coupling $J$, while they remain stable with respect to changes in $t$. The latter follows the opposite trend with $J$. Since the Kondo couplings are controlled by the gate voltages applied to the device in Fig.~\ref{fig:fig3} of the main text (gates $G_1$ and $G_2$), the results of Fig.~\ref{re_de}(c) are related to the gate-noise-induced dephasing rate.

For the relaxation, observing Fig.~\ref{re_de}(d), we note that there is no dependence in $J_{12}$, while the decay rate decreases with increasing $t$. Importantly, we also find that the qubit is fully protected against flux-noise-induced relaxation ($\bra{1}\frac{\partial H}{\partial \phi}\ket{0}=0$), since the qubit ground state is an eigenstate of the supercurrent operator across the junction, $\hat{\mathcal{J}} = \tfrac{\partial H}{\partial \phi}$ with eigenvalue zero for $\phi =0$. 

Finally, it is worth emphasizing that although the calculations above are performed for the microscopic fermionic model, the triple-quantum-dot device described in Fig.~\ref{fig:fig3} would ideally be embedded in a bosonic cQED architecture. This would enable coherent control of the device as well as fine-tuning against charge and flux noise~\cite{Gyenis2021}.

\section{Effect of Charge Fluctuations}\label{AppB}

\begin{figure}[b]
    \centering
\includegraphics[width=\columnwidth]{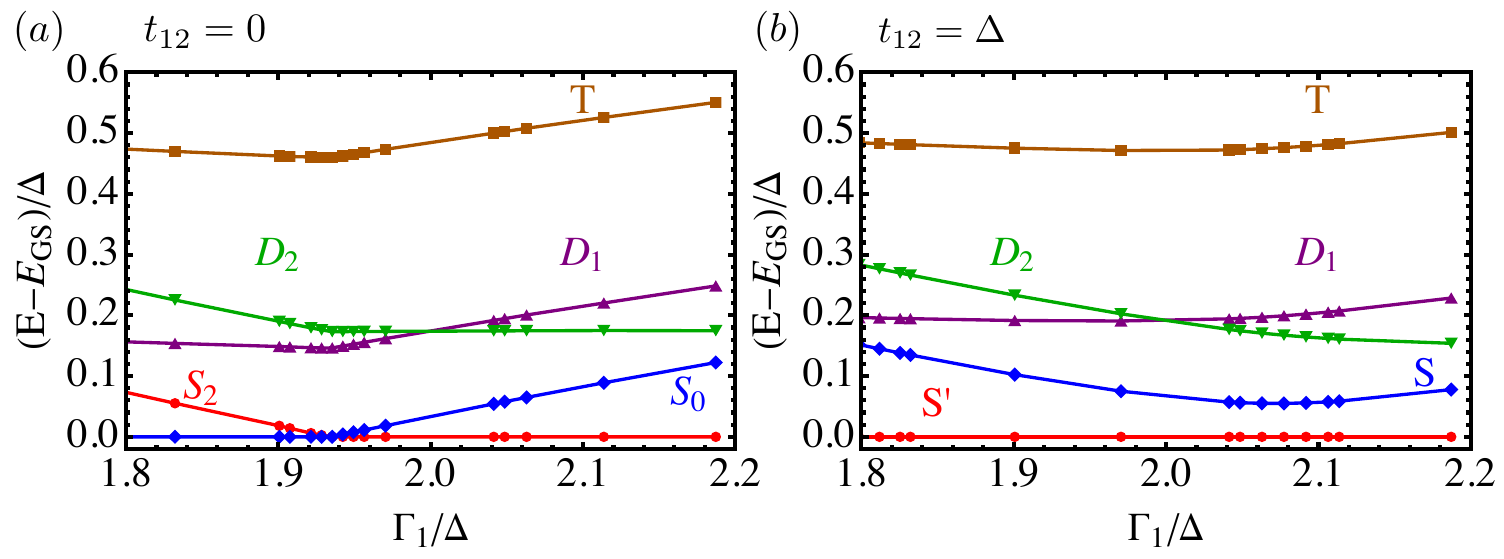}
    \caption{ 
   NRG results for the effective model (cf. Eq.~\eqref{eq:dots}) of a triple quantum dot device for $U_1=U_2=8\Delta, \epsilon=-0.5U,\Gamma_2=1.999\Delta, J_{12}=\Delta$ and $\Delta=D/400$ where $D$ is the band cut-off. $\Gamma_i=\pi t_i^2 \rho_0$ is the hopping rate.  The inter-bath hoppings are  (a) $t_{12}=0$ (b) $t_{12}=\Delta$. }\label{fig:2dot}
\end{figure}
％
In the main text, we have used an effective Kondo Hamiltonian to describe the coupling of the triple quantum dot system to the superconductors in a junction. In this section, 
we use the more fundamental Anderson model in order to assess the effect of charge fluctuations in the outer dots of the triple-dot system.  To this end, we study the following \emph{minimal} model of two Anderson impurities coupled to each other by super-exchange and tunnel coupled to two superconductors in a junction:
\begin{align}\label{eq:dots}
    H=& J_{12}   \bold{S}_1 \cdot \bold{S}_2  - t_{12} \sum_{\bold{k},\sigma} \left(c^\dag_{\bold{k}\sigma1}c_{\bold{k}\sigma 2}+ \mathrm{H.c.} \right) \notag\\
    &\qquad +\sum_{\alpha=1,2} H_\alpha \\
    H_\alpha =& U \left(n_{d;\uparrow\alpha}-\frac{1}{2}\right)\left(n_{d;\downarrow\alpha}-\frac{1}{2}\right)\notag\\
    &- t_\alpha \sum_{\bold{k} \sigma }\left(d^\dag_\alpha c_{\bold{k}\sigma\alpha}+ \mathrm{H.c.}\right) + \sum_{\bold{k},\sigma}  \epsilon_{\bold{k}} c^\dag_{\bold{k}\sigma\alpha}c_{\bold{k}\sigma\alpha} \notag\\
    &+\Delta \sum_{\bold{k} } \left( c_{\bold{k}\uparrow\alpha}c_{\bold{k}\downarrow\alpha}+ \mathrm{H.c.} \right).
\end{align}
Here $d_{\sigma\alpha}$ ($d^{\dag}_{\sigma\alpha}$)  is the annihilation  (creation) operator for electrons in the dot and $c_{\bold{k}\sigma\alpha}$
($c^{\dag}_{\bold{k}\sigma\alpha}$) is the annihilation (creation) operator 
for electrons in the superconductor $\alpha =1,2$ with s-wave pairing potential  proportional to $\Delta$. $U$ is the onsite coulomb interaction strength. We use the symmetric Anderson model to describe the dots. The global particle-hole symmetry is broken by the junction tunneling term $\propto t_{12}$.

Carrying out the calculation of the low-energy spectrum using NRG, we find a parameter regime accessible by tuning  the funneling into one of the  dots (which in turn controls the Kondo coupling) where the two lowest-energy states are spin-singlets and undergo an
avoided crossing, see
Fig.~\ref{fig:2dot}. Notice that, as shown in the figure,  the avoided crossing is observed for finite tunneling amplitude $t_{12}$ between the two superconductors in the junction (i.e. Panel (b) of the figure).

\section{On the choice of the Wilson chain}\label{AppC}

In order to carry out the Numerical Renormalization Group (NRG) calculations of the low-energy spectrum,  the Hamiltonian of the superconducting host (the bath) in Eq.~\eqref{eq:HeS} of the main text is discretized logarithmically using the adaptive scheme described in Ref.~\onlinecite{ZITKO20091271_discret} with discretization parameter $\Lambda=4$ assuming a constant density of states $\rho= 1/(2D)$. This procedure yields the following Wilson-chain Hamiltonian for the superconducting host:
\begin{align}\label{eq:wc}
H=&H_0 + \tilde{H}_{12}^X\\
H_0=&
H_{\text{imp}}+\tilde{H}_1+\tilde{H}_2,\\
H_\text{imp}=&J_{12} \bold{S_1}\cdot \bold{S_2}  +J_1 \bold{S}_1\cdot
\bold{s}_1+ J_2 \bold{S}_2\cdot\bold{s}_2.\\
\tilde{H}_\alpha=&V_\alpha n_{\alpha}(0) -\sum_i t_{i}\left[f_{\alpha\sigma}(i)f^\dag_{\alpha\sigma}(i+1)+ \mathrm{H.c.} \right]\notag\\
&+\Delta\left[f_{\alpha\uparrow}(i)f_{\alpha\downarrow}(i)+ \mathrm{H.c.}\right],
\end{align}
where $t_i\sim \Lambda^{-i/2}$ decays exponentially. We note that the Hamiltonian, $H_0$, for $V_\alpha\neq0$ only preserves the SU(2) spin symmetry. However, 
 if $V_\alpha=0$, there is a hidden U(1) charge symmetry~\cite{PhysRevB.90.241108_YAO,PhysRevB.78.245109_symmetry}, whose generator is $Q_{x;\alpha} = (Q_{+;\alpha}+Q_{-;\alpha})/2$, where 
 \begin{align}
 Q_{+;\alpha} &= \sum_{i} (-1)^i f^\dag_{\alpha\uparrow}(i)f^\dag_{\alpha\downarrow}(i), \\
Q_{-;\alpha} &= Q_{+;\alpha}^\dag.
 \end{align}
This U($1$) quantum number plays an essential role in our qubit systems.

 $\tilde{H}^X_{12}$ denotes the term in the Hamiltonian  coupling the two superconducting baths (channels). For the TSC-superconductor system, the tunneling reads, 
\begin{align}
\tilde{H}^\text{TSC}_{12}&=-t_{i}\mathcal{S}_{12} \sum_{i\sigma}  \left[f_{1\sigma}(i)f^\dag_{2\sigma}(i+1)\right.\notag\\
&\left.+f_{2\sigma}(i)f^\dag_{1\sigma}(i+1)+ \mathrm{H.c.} \right],
\end{align}
where $\mathcal{S}_{12} = \mathcal{S}(\epsilon_F)$ is the tunneling amplitude for the electrons at the Fermi level to propagate from impurity $1$ to impurity $2$~\cite{PhysRevB.90.241108_YAO}. This Hamiltonian preserves the pseudo-charge quantum number, $Q^+_x=\sum_\alpha Q_{x;\alpha}$, i.e. $[Q^+_x,\tilde{H}^{TSC}_{12}]=0$. Furthermore, the two singlets have different quantum number $Q_x^+$. To generate an avoided crossing, we need to break this symmetry by including the scattering potentials that couple to the local densities at the first sites of the channel Wilson chains.


On the other hand, for the quantum-dot device, the tunneling reads,
\begin{align}
\tilde{H}^\text{QD}_{12}&=-t_{12} \sum_{i\sigma}  \left[f_{1\sigma}(i)f^\dag_{2\sigma}(i)+f_{2\sigma}(i)f^\dag_{1\sigma}(i) \right].
\end{align}
In this case,  the conserved pseudo-charge generator is not $Q^{+}_x$ but $Q^-_x=\sum_\alpha Q_{x;\alpha}(-1)^\alpha$.  The two singlets have the same $Q^-_x$, which allows the hopping term to mix the singlet states and leads to an avoided crossing. 
The difference between these two models (TSC and quantum-dot device) can be understood by considering the particle-hole transformations generated by $Q^{\pm}_{x}$:
\begin{align}
f_{\alpha\sigma}(i) \to P(i,\alpha) f^\dag_{\alpha,-\sigma}(i)\\
f_{\alpha\sigma}^\dag(i) \to P(i,\alpha) f_{\alpha,-\sigma}(i)
\end{align}
where $P(i,\alpha)=(-1)^{i}$  as generated by $Q^{+}_x$ for the TSC system, and $P(i,\alpha)=(-1)^{i+\alpha}$, as generated by $Q^{-}_x$ for quantum-dot device. That is, the two models are invariant under different particle-hole transformations. For the TSC system with $V_1=V_2=0$, the two Wilson chains undergo the same  (global) particle-hole transformation since they are derived from the Hamiltonian of a single superconductor~\cite{yao2014}. On the contrary, the Wilson chains for the quantum dot device describe  two independent superconductors that are coupled by tunneling. As it has been mentioned in the main text and above in the discussion of the ZBA, the latter breaks the global particle-hole symmetry generated by $Q^{+}_x$.
\section{Details of the NRG  for the device}\label{AppD}
For the calculation of the low-energy spectrum and the time evolution of the QD device, we set $V_{\alpha}=0$. In this limit, we   further simplify  the Hamiltonian following the method described in \cite{JPSJ.67.1332_BCS_trans} which corresponds to a rotation in the conserved $Q_x^-$ sector. We apply the  following Bogoliubov transformation:
\begin{align}
b^\dag_{\alpha\uparrow}(i)&=\frac{1}{\sqrt{2}} \left(f^\dag_{\alpha\uparrow}(i)+f_{\alpha\downarrow}(i)  \right),\\
b_{\alpha\downarrow}(i)&=\frac{1}{\sqrt{2}} \left(f^\dag_{\alpha\uparrow}(i)- f_{\alpha\downarrow}(i)  \right),
\end{align}
followed by a particle-hole transformation for chain 1:
\begin{align}
c^\dag_{1\uparrow}(2i)&=b^\dag_{1\uparrow}(2i),\\
c_{1\downarrow}(2i)&= b_{1\downarrow}(2i),\\
c^\dag_{1\uparrow}(2i-1)&=b_{1\downarrow}(2i-1),\\
c_{1\downarrow}(2i-1)&=-b^\dag_{1\uparrow}(2i-1).
\end{align}
For chain 2:
\begin{align}
c^\dag_{2\uparrow}(2i-1)&=b^\dag_{2\uparrow}(2i-1),\\
c_{2\downarrow}(2i-1)&= b_{2\downarrow}(2i-1),\\
c^\dag_{2\uparrow}(2i)&=b_{2\downarrow}(2i),\\
c_{2\downarrow}(2i)&=-b^\dag_{2\uparrow}(2i).
\end{align}
which yields the following Wilson-chain Hamiltonian:
\begin{align}
H&=H_{\text{imp}}+\tilde{H}'_1+\tilde{H}'_2+\tilde{H}^{'QD}_{12},\\
\tilde{H}'_\alpha&=\sum_i \left\{ t_{i}\sum_\sigma \left( c_{\alpha\sigma}(i)c_{\alpha\sigma}^\dag (i+1)+h.c. \right)\right.\notag\\
&-  \left.\Delta (-1)^{i+\alpha} Q_{Z;\alpha}(i)   \right\}.
\end{align}
and
\begin{align}
\tilde{H}^{'QD}_{12}& = -t_{12}\sum_{\sigma} \left[  c_{2\sigma}(i)c_{1\sigma}^\dag(i)+c_{1\sigma}(i)c^\dag_{2\sigma}(i) \right].
\end{align}
 The NRG for the model in Eq.~\eqref{eq:HeS} with constant overlap matrix $\mathcal{S}_{12}$ (corresponding to the TSC-superconductor system) is performed using the conserved SU($2$) spin, $S$. For the device,  we used both the SU($2$) spin and U($1$) pseudo-charge quantum numbers $(S,Q_{Z})$:
\begin{align}
Q_{Z}& = Q_{Z;1}+Q_{Z;2},\\
Q_{Z;\alpha}&=\sum_{i} Q_{Z;\alpha}(i) =\frac{1}{2} \sum_i \left[ n_{\alpha\uparrow}(i)+n_{\alpha\downarrow}(i)-1 \right]. \\
S&=\sum_\alpha \left[S_{\alpha}^\text{imp} + \sum_i S_{\alpha}(i) \right].
\end{align}
We keep at most $5,000$ multiplets ($\sim 15,000$ states) for the spectrum and evolution of the device. In the presence of a gap, the NRG iteration should be truncated at iterations with energy scale $\omega_N\ll \Delta$~\cite{Hecht_2008_BCS}. We stop our NRG computation at iterations with energy scale $\sim 10^{-9}\Delta$ and use a temperature $T\ll\Delta$ as an effective zero-temperature limit.  
\section{Time-dependent NRG calculations}\label{AppE} 

In this section, we provide the details of the time-dependent NRG calculation for the overlap after a multiple quench. For the quench profiles of the series of Hamiltonians $H_0, H_1, \cdots, H_n$ with periods $t_0=0, t_1, \cdots, t_n$ starting from initial states $| l_0e^N_0N\rangle$ and $| l'_0e^N_0N\rangle$ where $|l_i e_i N \rangle $ denotes the state of Hamiltonian $H_i$ at the last iteration $N$ labeled by $l_i$ and the environment index $e_i$. By the definition, within the Full-Density Matrix NRG  (FDM-NRG)~\cite{PhysRevLett.99.076402_FDM}, all states are discarded at the last iteration. The overlap at time $t =t_1+t_2+\cdots+t_n$ reads
\begin{align}
P_{l_0l_0'}(t)= \langle  l'_0 e^N_0 N | e^{-i H_n t_n} \cdots e^{-i H_1 t_1} |l_0e^N_0N \rangle.
\end{align}
Next, we insert the complete basis~\cite{AS_PhysRevLett.95.196801,td_NRG_Smatrix_PhysRevB.74.245113} for each NRG of Hamiltonian $H_i$:
\begin{align}
1_i&=\sum_{l_ie_im_i} | l_ie_im_i\rangle  \: \langle l_ie_im_i|,\label{eq:AS_1}\\
&=\sum_{l_ie_im_i<m'_i} | l_ie_im_i\rangle  \: \langle l_ie_im_i| + \sum_{s_i e_i} | s_ie_im'_i\rangle  \: \langle s_ie_im'_i|. \label{eq:AS_2}
\end{align}
where $l_i$ are the labels for all the discarded states at iteration $m_i$ and $s_i$ labels both the discarded and kept states. $e_i$ is the index for the environment basis. Both of the identities will be used in the following derivations. Using the first equation, we get,

\begin{widetext}
\begin{align}
&P_{l_0l_0'}=\sum_{l_n,e_n,m_n}\cdots \sum_{l_2,e_2,m_2}\sum_{l_1,e_1,m_1}\:\langle l'_0e^N_0N |  l_ne_nm_n\rangle  \: \langle l_ne_nm_n|e^{-i H_n t_n} \cdots \notag\\
&| l_2e_2m_2\rangle  \: \langle l_2e_2m_2| e^{-iH_2 t_2} | l_1e_1m_1\rangle  \: \langle l_1e_1m_1| e^{-i H_1 t_1} |l_0e^N_0N \rangle.
\end{align}
\end{widetext}

Using the NRG approximation $ \langle l_ie_im_i| e^{-iH_i t_i}  = \langle l_ie_im_i| e^{-i\epsilon(l_i,m_i) t_i}$ where $\epsilon(l_i,m_i)$ is the eigenvalue of the eigenstate $\:\langle l_ie_im_i|$ and the following identity for the multiple sums~\cite{FDM_identity_PhysRevB.92.155435}:
\begin{align}
\sum_{l_n,e_n,m_n}\cdots \sum_{l_2,e_2,m_2}\sum_{l_1,e_1,m_1} = \sum^{\neq K_1,K_2,\cdots,K_n}_{r_1,r_2,\cdots,r_n}  \sum_{e_1,e_2,\cdots,e_n} \sum_m
\end{align}
In the right-hand side, $r_1,\cdots,r_n$ label both kept or discarded states, and they cannot be all kept states in the summation. We have
\begin{widetext}
\begin{equation}
\begin{split}
P_{l_0l_0'}=& \sum^{\neq K_1,K_2,\cdots,K_n}_{r_1,r_2,\cdots,r_n}  \sum_{e_1,e_2,\cdots,e_N} \sum_m \langle l'_0e^N_0N |  r_ne_nm\rangle  \: \langle r_ne_nm|r_{n-1}e_{n-1}m\rangle e^{-i \epsilon(r_n,m) t_n} \cdots\notag,\\
&\times| r_2e_2m\rangle  \: e^{-i \epsilon(r_2,m) t_2} \langle r_2e_2m | r_1e_1m\rangle  e^{-i \epsilon(r_1,m) t_1}  \langle r_1e_1m |l_0e^N_0N \rangle,\\
&= \sum^{\neq K_1,K_2,\cdots,K_n}_{r_1,r_2,\cdots,r_n}  \sum_{e_1,e_2,\cdots,e_n} \sum_m \langle l'_0e^N_0N |  r_ne_nm\rangle  \:  S_{r_n,r_{n-1}}(m)\delta_{e_n,e_{n-1}} e^{-i \epsilon(r_n,m) t_n} \cdots\notag\\
&  \times e^{-i \epsilon(r_2,m) t_2}  S_{r_2,r_{1}}(m)\delta_{e_2,e_{1}}   e^{-i \epsilon(r_1,m) t_1}  \langle r_1e_1m |l_0e^N_0N \rangle,\\
=& \sum^{\neq K_1,K_2,\cdots,K_n}_{r_1,r_2,\cdots,r_n}  \sum_{e} \sum_m \langle r_1e m |l_0e^N_0N \rangle \:\langle l'_0e^N_0N |  r_ne m\rangle  \:  \times\\ \notag
& S_{r_n,r_{n-1}}(m)  e^{-i \epsilon(r_n,m) t_n} \cdots\notag e^{-i \epsilon(r_2,m) t_2}  S_{r_2,r_{1}}(m)    e^{-i \epsilon(r_1,m) t_1} \label{eq:P1},
\end{split}
\end{equation}
\end{widetext}
where $S_{r_nr_{n-1}}(m) \delta_{e_n,e_{n-1}} = \langle r_n e_n m | r_{n-1} e_{n-1} m \rangle$ is the overlap matrix element between NRG eigenstates of different Hamiltonians~\cite{td_NRG_Smatrix_PhysRevB.74.245113}. Next, we calculate the following matrix element by inserting the second identity in Eq.~\eqref{eq:AS_2}
\begin{widetext}
\begin{align}
&\sum_{e}  \langle r_1e m |l_0e^N_0N \rangle \:\langle l'_0e^N_0N |  r_ne m\rangle,\\
&=\sum_{e}  \langle r_1e m |\left[\sum_{l_0e_0m_0<m} | l_0e_0m_0\rangle  \: \langle l_0e_0m_0| + \sum_{s_0 e_0} | s_0e_0m\rangle  \: \langle s_0e_0m| \right]|l_0e^N_0N \rangle \notag\\
&\times\:\langle l'_0e^N_0N |\left[\sum_{l_0e_0m_0<m} | l_0e_0m_0\rangle  \: \langle l_0e_0m_0| + \sum_{s_0 e_0} | s_0e_0m\rangle  \: \langle s_0e_0m| \right]|  r_ne m\rangle\\
&=\sum_{e} \sum_{s_0 e_0} \sum_{s'_0 e'_0} \langle r_1e m |  s_0e_0m\rangle  \: \langle s_0e_0m| l_0e^N_0N \rangle \:\langle l'_0e^N_0N |     s'_0e'_0m\rangle  \: \langle s'_0e'_0m |  r_ne m\rangle\\
&=\sum_{e} \sum_{s_0 e_0} \sum_{s'_0 e'_0}S_{r_1s_0}(m) \delta_{e,e_0}  \langle s_0e_0m| l_0e^N_0N \rangle \:\langle l'_0e^N_0N |     s'_0e'_0m\rangle  S_{s'_0r_n}(m) \delta_{e'_0,e} \\
&= \sum_{s_0  s'_0  }S_{r_1s_0}(m) \left[ \sum_{e}  \left[A^{N\to m}_{l_0e^N_0\to s_0e} \right]^\dag O_{l_0l_0'}(N) A^{N\to m}_{l'_0e_0^N\to s'_0e} \right]S_{s'_0r_n}(m)  \\
&= \sum_{s_0  s'_0  }S_{r_1s_0}(m)  O^{red}_{l_0l'_0\to s_0 s'_0}(m) S_{s'_0r_n}(m) \label{eq:rho_red}
\end{align}
\end{widetext}
where $O_{l_0l_0'}(N) = | l_0e^N_0N \rangle \:\langle l'_0e^N_0N |$ is the matrix element at the last iteration of NRG on $H_0$ and $O^{red}_{l_0l'_0\to s_0 s'_0}(m)$ is the reduced matrix element which is calculated by the tensor product of the unitary transformations $A^{n\to n-1}_{le\to l'e'}$ calculated from each NRG iteration of $H_0$ and tracing out the environment basis~\cite{PhysRevLett.99.076402_FDM}. Combining Eq.~\eqref{eq:P1} and \eqref{eq:rho_red}. The overlap is
\begin{align}\label{eq:P_fin}
P_{l_0l_0'}=&\sum^{\neq K_1,K_2,\cdots,K_n}_{r_1,r_2,\cdots,r_n}    \sum_m \sum_{s_0  s'_0  }  e^{-i \epsilon(r_n,m) t_n}  S_{r_n,r_{n-1}}(m)   \times\cdots \\ \notag
&\times    e^{-i \epsilon(r_1,m) t_1}S_{r_1s_0}(m)  O^{red}_{l_0l'_0\to s_0 s'_0}(m) S_{s'_0r_n}(m).
\end{align}
To compute this, we first compute all the NRG eigenvectors and eigenvalues for the Hamiltonians, $H_0,\cdots,H_n$. Then use the eigenvectors to compute all the necessary overlap matrix element, $S_{ij}$ and the reduced matrix elements, $O^{red}$. Finally inserting everything into Eq.~\eqref{eq:P_fin} leads to the result.

\nocite{apsrev42Control}
\bibliographystyle{apsrev4-2} 
\bibliography{biblio_all}

\begin{thebibliography}{81}%
\makeatletter
\providecommand \@ifxundefined [1]{%
 \@ifx{#1\undefined}
}%
\providecommand \@ifnum [1]{%
 \ifnum #1\expandafter \@firstoftwo
 \else \expandafter \@secondoftwo
 \fi
}%
\providecommand \@ifx [1]{%
 \ifx #1\expandafter \@firstoftwo
 \else \expandafter \@secondoftwo
 \fi
}%
\providecommand \natexlab [1]{#1}%
\providecommand \enquote  [1]{``#1''}%
\providecommand \bibnamefont  [1]{#1}%
\providecommand \bibfnamefont [1]{#1}%
\providecommand \citenamefont [1]{#1}%
\providecommand \href@noop [0]{\@secondoftwo}%
\providecommand \href [0]{\begingroup \@sanitize@url \@href}%
\providecommand \@href[1]{\@@startlink{#1}\@@href}%
\providecommand \@@href[1]{\endgroup#1\@@endlink}%
\providecommand \@sanitize@url [0]{\catcode `\\12\catcode `\$12\catcode
  `\&12\catcode `\#12\catcode `\^12\catcode `\_12\catcode `\%12\relax}%
\providecommand \@@startlink[1]{}%
\providecommand \@@endlink[0]{}%
\providecommand \url  [0]{\begingroup\@sanitize@url \@url }%
\providecommand \@url [1]{\endgroup\@href {#1}{\urlprefix }}%
\providecommand \urlprefix  [0]{URL }%
\providecommand \Eprint [0]{\href }%
\providecommand \doibase [0]{https://doi.org/}%
\providecommand \selectlanguage [0]{\@gobble}%
\providecommand \bibinfo  [0]{\@secondoftwo}%
\providecommand \bibfield  [0]{\@secondoftwo}%
\providecommand \translation [1]{[#1]}%
\providecommand \BibitemOpen [0]{}%
\providecommand \bibitemStop [0]{}%
\providecommand \bibitemNoStop [0]{.\EOS\space}%
\providecommand \EOS [0]{\spacefactor3000\relax}%
\providecommand \BibitemShut  [1]{\csname bibitem#1\endcsname}%
\let\auto@bib@innerbib\@empty
\bibitem [{\citenamefont {Giamarchi}(2003)}]{Giamarchi2003}%
  \BibitemOpen
  \bibfield  {author} {\bibinfo {author} {\bibfnamefont {T.}~\bibnamefont
  {Giamarchi}},\ }\href
  {https://doi.org/10.1093/acprof:oso/9780198525004.001.0001} {\emph {\bibinfo
  {title} {Quantum Physics in One Dimension}}}\ (\bibinfo  {publisher} {Oxford
  University Press},\ \bibinfo {year} {2003})\BibitemShut {NoStop}%
\bibitem [{\citenamefont {Haldane}(1983{\natexlab{a}})}]{Haldane1983}%
  \BibitemOpen
  \bibfield  {author} {\bibinfo {author} {\bibfnamefont {F.~D.~M.}\
  \bibnamefont {Haldane}},\ }\bibfield  {title} {\bibinfo {title} {Nonlinear
  field theory of large-spin heisenberg antiferromagnets: Semiclassically
  quantized solitons of the one-dimensional easy-axis n\'eel state},\ }\href
  {https://doi.org/10.1103/PhysRevLett.50.1153} {\bibfield  {journal} {\bibinfo
   {journal} {Phys. Rev. Lett.}\ }\textbf {\bibinfo {volume} {50}},\ \bibinfo
  {pages} {1153} (\bibinfo {year} {1983}{\natexlab{a}})}\BibitemShut {NoStop}%
\bibitem [{\citenamefont {Haldane}(1983{\natexlab{b}})}]{HALDANE1983464}%
  \BibitemOpen
  \bibfield  {author} {\bibinfo {author} {\bibfnamefont {F.}~\bibnamefont
  {Haldane}},\ }\bibfield  {title} {\bibinfo {title} {Continuum dynamics of the
  1-d heisenberg antiferromagnet: Identification with the o(3) nonlinear sigma
  model},\ }\href
  {https://doi.org/https://doi.org/10.1016/0375-9601(83)90631-X} {\bibfield
  {journal} {\bibinfo  {journal} {Physics Letters A}\ }\textbf {\bibinfo
  {volume} {93}},\ \bibinfo {pages} {464} (\bibinfo {year}
  {1983}{\natexlab{b}})}\BibitemShut {NoStop}%
\bibitem [{\citenamefont {Affleck}\ \emph {et~al.}(1987)\citenamefont
  {Affleck}, \citenamefont {Kennedy}, \citenamefont {Lieb},\ and\ \citenamefont
  {Tasaki}}]{PhysRevLett.59.799}%
  \BibitemOpen
  \bibfield  {author} {\bibinfo {author} {\bibfnamefont {I.}~\bibnamefont
  {Affleck}}, \bibinfo {author} {\bibfnamefont {T.}~\bibnamefont {Kennedy}},
  \bibinfo {author} {\bibfnamefont {E.~H.}\ \bibnamefont {Lieb}},\ and\
  \bibinfo {author} {\bibfnamefont {H.}~\bibnamefont {Tasaki}},\ }\bibfield
  {title} {\bibinfo {title} {Rigorous results on valence-bond ground states in
  antiferromagnets},\ }\href {https://doi.org/10.1103/PhysRevLett.59.799}
  {\bibfield  {journal} {\bibinfo  {journal} {Phys. Rev. Lett.}\ }\textbf
  {\bibinfo {volume} {59}},\ \bibinfo {pages} {799} (\bibinfo {year}
  {1987})}\BibitemShut {NoStop}%
\bibitem [{\citenamefont {Hagiwara}\ \emph
  {et~al.}(1990{\natexlab{a}})\citenamefont {Hagiwara}, \citenamefont
  {Katsumata}, \citenamefont {Affleck}, \citenamefont {Halperin},\ and\
  \citenamefont {Renard}}]{Affleck1990}%
  \BibitemOpen
  \bibfield  {author} {\bibinfo {author} {\bibfnamefont {M.}~\bibnamefont
  {Hagiwara}}, \bibinfo {author} {\bibfnamefont {K.}~\bibnamefont {Katsumata}},
  \bibinfo {author} {\bibfnamefont {I.}~\bibnamefont {Affleck}}, \bibinfo
  {author} {\bibfnamefont {B.~I.}\ \bibnamefont {Halperin}},\ and\ \bibinfo
  {author} {\bibfnamefont {J.~P.}\ \bibnamefont {Renard}},\ }\bibfield  {title}
  {\bibinfo {title} {Observation of s=1/2 degrees of freedom in an s=1
  linear-chain heisenberg antiferromagnet},\ }\href
  {https://doi.org/10.1103/PhysRevLett.65.3181} {\bibfield  {journal} {\bibinfo
   {journal} {Phys. Rev. Lett.}\ }\textbf {\bibinfo {volume} {65}},\ \bibinfo
  {pages} {3181} (\bibinfo {year} {1990}{\natexlab{a}})}\BibitemShut {NoStop}%
\bibitem [{\citenamefont {Kitaev}(2001)}]{Kitaev2001}%
  \BibitemOpen
  \bibfield  {author} {\bibinfo {author} {\bibfnamefont {A.~Y.}\ \bibnamefont
  {Kitaev}},\ }\bibfield  {title} {\bibinfo {title} {Unpaired majorana fermions
  in quantum wires},\ }\href {https://doi.org/10.1070/1063-7869/44/10S/S29}
  {\bibfield  {journal} {\bibinfo  {journal} {Physics-Uspekhi}\ }\textbf
  {\bibinfo {volume} {44}},\ \bibinfo {pages} {131} (\bibinfo {year}
  {2001})}\BibitemShut {NoStop}%
\bibitem [{\citenamefont {Pollmann}\ \emph {et~al.}(2012)\citenamefont
  {Pollmann}, \citenamefont {Berg}, \citenamefont {Turner},\ and\ \citenamefont
  {Oshikawa}}]{Pollmann2012}%
  \BibitemOpen
  \bibfield  {author} {\bibinfo {author} {\bibfnamefont {F.}~\bibnamefont
  {Pollmann}}, \bibinfo {author} {\bibfnamefont {E.}~\bibnamefont {Berg}},
  \bibinfo {author} {\bibfnamefont {A.~M.}\ \bibnamefont {Turner}},\ and\
  \bibinfo {author} {\bibfnamefont {M.}~\bibnamefont {Oshikawa}},\ }\bibfield
  {title} {\bibinfo {title} {Symmetry protection of topological phases in
  one-dimensional quantum spin systems},\ }\href
  {https://doi.org/10.1103/PhysRevB.85.075125} {\bibfield  {journal} {\bibinfo
  {journal} {Phys. Rev. B}\ }\textbf {\bibinfo {volume} {85}},\ \bibinfo
  {pages} {075125} (\bibinfo {year} {2012})}\BibitemShut {NoStop}%
\bibitem [{\citenamefont {Nadj-Perge}\ \emph {et~al.}(2013)\citenamefont
  {Nadj-Perge}, \citenamefont {Drozdov}, \citenamefont {Bernevig},\ and\
  \citenamefont {Yazdani}}]{Bernevig2013}%
  \BibitemOpen
  \bibfield  {author} {\bibinfo {author} {\bibfnamefont {S.}~\bibnamefont
  {Nadj-Perge}}, \bibinfo {author} {\bibfnamefont {I.~K.}\ \bibnamefont
  {Drozdov}}, \bibinfo {author} {\bibfnamefont {B.~A.}\ \bibnamefont
  {Bernevig}},\ and\ \bibinfo {author} {\bibfnamefont {A.}~\bibnamefont
  {Yazdani}},\ }\bibfield  {title} {\bibinfo {title} {Proposal for realizing
  majorana fermions in chains of magnetic atoms on a superconductor},\ }\href
  {https://doi.org/10.1103/PhysRevB.88.020407} {\bibfield  {journal} {\bibinfo
  {journal} {Phys. Rev. B}\ }\textbf {\bibinfo {volume} {88}},\ \bibinfo
  {pages} {020407} (\bibinfo {year} {2013})}\BibitemShut {NoStop}%
\bibitem [{\citenamefont {Nadj-Perge}\ \emph {et~al.}(2014)\citenamefont
  {Nadj-Perge}, \citenamefont {Drozdov}, \citenamefont {Li}, \citenamefont
  {Chen}, \citenamefont {Jeon}, \citenamefont {Seo}, \citenamefont {MacDonald},
  \citenamefont {Bernevig},\ and\ \citenamefont {Yazdani}}]{Yazdani2014}%
  \BibitemOpen
  \bibfield  {author} {\bibinfo {author} {\bibfnamefont {S.}~\bibnamefont
  {Nadj-Perge}}, \bibinfo {author} {\bibfnamefont {I.}~\bibnamefont {Drozdov}},
  \bibinfo {author} {\bibfnamefont {J.}~\bibnamefont {Li}}, \bibinfo {author}
  {\bibfnamefont {H.}~\bibnamefont {Chen}}, \bibinfo {author} {\bibfnamefont
  {S.}~\bibnamefont {Jeon}}, \bibinfo {author} {\bibfnamefont {J.}~\bibnamefont
  {Seo}}, \bibinfo {author} {\bibfnamefont {A.}~\bibnamefont {MacDonald}},
  \bibinfo {author} {\bibfnamefont {B.}~\bibnamefont {Bernevig}},\ and\
  \bibinfo {author} {\bibfnamefont {A.}~\bibnamefont {Yazdani}},\ }\bibfield
  {title} {\bibinfo {title} {Observation of majorana fermions in ferromagnetic
  atomic chains on a superconductor},\ }\href
  {https://doi.org/10.1126/science.1259327} {\bibfield  {journal} {\bibinfo
  {journal} {Science}\ }\textbf {\bibinfo {volume} {346}},\ \bibinfo {pages}
  {602} (\bibinfo {year} {2014})}\BibitemShut {NoStop}%
\bibitem [{\citenamefont {Grill}\ \emph {et~al.}(2007)\citenamefont {Grill},
  \citenamefont {Dyer}, \citenamefont {Lafferentz}, \citenamefont {Persson},
  \citenamefont {Peters},\ and\ \citenamefont {Hecht}}]{grill2007}%
  \BibitemOpen
  \bibfield  {author} {\bibinfo {author} {\bibfnamefont {L.}~\bibnamefont
  {Grill}}, \bibinfo {author} {\bibfnamefont {M.}~\bibnamefont {Dyer}},
  \bibinfo {author} {\bibfnamefont {L.}~\bibnamefont {Lafferentz}}, \bibinfo
  {author} {\bibfnamefont {M.}~\bibnamefont {Persson}}, \bibinfo {author}
  {\bibfnamefont {M.~V.}\ \bibnamefont {Peters}},\ and\ \bibinfo {author}
  {\bibfnamefont {S.}~\bibnamefont {Hecht}},\ }\bibfield  {title} {\bibinfo
  {title} {Nano-architectures by covalent assembly of molecular building
  blocks},\ }\href {https://doi.org/10.1038/nnano.2007.346} {\bibfield
  {journal} {\bibinfo  {journal} {Nature Nanotech}\ }\textbf {\bibinfo {volume}
  {2}},\ \bibinfo {pages} {687} (\bibinfo {year} {2007})}\BibitemShut {NoStop}%
\bibitem [{\citenamefont {Gourdon}(2008)}]{gourdon2008}%
  \BibitemOpen
  \bibfield  {author} {\bibinfo {author} {\bibfnamefont {A.}~\bibnamefont
  {Gourdon}},\ }\bibfield  {title} {\bibinfo {title} {On‐{{Surface Covalent
  Coupling}} in {{Ultrahigh Vacuum}}},\ }\href
  {https://doi.org/10.1002/anie.200802229} {\bibfield  {journal} {\bibinfo
  {journal} {Angewandte Chemie}\ }\textbf {\bibinfo {volume} {47}},\ \bibinfo
  {pages} {6950} (\bibinfo {year} {2008})}\BibitemShut {NoStop}%
\bibitem [{\citenamefont {Pavliček}\ \emph {et~al.}(2017)\citenamefont
  {Pavliček}, \citenamefont {Mistry}, \citenamefont {Majzik}, \citenamefont
  {Moll}, \citenamefont {Meyer}, \citenamefont {Fox},\ and\ \citenamefont
  {Gross}}]{pavlicek2017}%
  \BibitemOpen
  \bibfield  {author} {\bibinfo {author} {\bibfnamefont {N.}~\bibnamefont
  {Pavliček}}, \bibinfo {author} {\bibfnamefont {A.}~\bibnamefont {Mistry}},
  \bibinfo {author} {\bibfnamefont {Z.}~\bibnamefont {Majzik}}, \bibinfo
  {author} {\bibfnamefont {N.}~\bibnamefont {Moll}}, \bibinfo {author}
  {\bibfnamefont {G.}~\bibnamefont {Meyer}}, \bibinfo {author} {\bibfnamefont
  {D.~J.}\ \bibnamefont {Fox}},\ and\ \bibinfo {author} {\bibfnamefont
  {L.}~\bibnamefont {Gross}},\ }\bibfield  {title} {\bibinfo {title} {Synthesis
  and characterization of triangulene},\ }\href
  {https://doi.org/10.1038/nnano.2016.305} {\bibfield  {journal} {\bibinfo
  {journal} {Nature Nanotech}\ }\textbf {\bibinfo {volume} {12}},\ \bibinfo
  {pages} {308} (\bibinfo {year} {2017})}\BibitemShut {NoStop}%
\bibitem [{\citenamefont {Mishra}\ \emph {et~al.}(2019)\citenamefont {Mishra},
  \citenamefont {Beyer}, \citenamefont {Eimre}, \citenamefont {Liu},
  \citenamefont {Berger}, \citenamefont {Gr{\"o}ning}, \citenamefont
  {Pignedoli}, \citenamefont {M{\"u}llen}, \citenamefont {Fasel}, \citenamefont
  {Feng},\ and\ \citenamefont {Ruffieux}}]{Mishra2019}%
  \BibitemOpen
  \bibfield  {author} {\bibinfo {author} {\bibfnamefont {S.}~\bibnamefont
  {Mishra}}, \bibinfo {author} {\bibfnamefont {D.}~\bibnamefont {Beyer}},
  \bibinfo {author} {\bibfnamefont {K.}~\bibnamefont {Eimre}}, \bibinfo
  {author} {\bibfnamefont {J.}~\bibnamefont {Liu}}, \bibinfo {author}
  {\bibfnamefont {R.}~\bibnamefont {Berger}}, \bibinfo {author} {\bibfnamefont
  {O.}~\bibnamefont {Gr{\"o}ning}}, \bibinfo {author} {\bibfnamefont {C.~A.}\
  \bibnamefont {Pignedoli}}, \bibinfo {author} {\bibfnamefont {K.}~\bibnamefont
  {M{\"u}llen}}, \bibinfo {author} {\bibfnamefont {R.}~\bibnamefont {Fasel}},
  \bibinfo {author} {\bibfnamefont {X.}~\bibnamefont {Feng}},\ and\ \bibinfo
  {author} {\bibfnamefont {P.}~\bibnamefont {Ruffieux}},\ }\bibfield  {title}
  {\bibinfo {title} {Synthesis and characterization of $\pi$-extended
  triangulene},\ }\href {https://doi.org/10.1021/jacs.9b05319} {\bibfield
  {journal} {\bibinfo  {journal} {Journal of the American Chemical Society}\
  }\textbf {\bibinfo {volume} {141}},\ \bibinfo {pages} {10621} (\bibinfo
  {year} {2019})}\BibitemShut {NoStop}%
\bibitem [{\citenamefont {Vilas-Varela}\ \emph {et~al.}(2023)\citenamefont
  {Vilas-Varela}, \citenamefont {Romero-Lara}, \citenamefont {Vegliante},
  \citenamefont {Calupitan}, \citenamefont {Mart{\'\i}nez}, \citenamefont
  {Meyer}, \citenamefont {Uriarte-Amiano}, \citenamefont {Friedrich},
  \citenamefont {Wang}, \citenamefont {Schulz} \emph
  {et~al.}}]{vilas2023surface}%
  \BibitemOpen
  \bibfield  {author} {\bibinfo {author} {\bibfnamefont {M.}~\bibnamefont
  {Vilas-Varela}}, \bibinfo {author} {\bibfnamefont {F.}~\bibnamefont
  {Romero-Lara}}, \bibinfo {author} {\bibfnamefont {A.}~\bibnamefont
  {Vegliante}}, \bibinfo {author} {\bibfnamefont {J.~P.}\ \bibnamefont
  {Calupitan}}, \bibinfo {author} {\bibfnamefont {A.}~\bibnamefont
  {Mart{\'\i}nez}}, \bibinfo {author} {\bibfnamefont {L.}~\bibnamefont
  {Meyer}}, \bibinfo {author} {\bibfnamefont {U.}~\bibnamefont
  {Uriarte-Amiano}}, \bibinfo {author} {\bibfnamefont {N.}~\bibnamefont
  {Friedrich}}, \bibinfo {author} {\bibfnamefont {D.}~\bibnamefont {Wang}},
  \bibinfo {author} {\bibfnamefont {F.}~\bibnamefont {Schulz}}, \emph
  {et~al.},\ }\bibfield  {title} {\bibinfo {title} {On-surface synthesis and
  characterization of a high-spin aza-[5]-triangulene},\ }\href
  {https://doi.org/10.1002/anie.202307884} {\bibfield  {journal} {\bibinfo
  {journal} {Angewandte Chemie}\ }\textbf {\bibinfo {volume} {135}},\ \bibinfo
  {pages} {e202307884} (\bibinfo {year} {2023})}\BibitemShut {NoStop}%
\bibitem [{\citenamefont {Mishra}\ \emph {et~al.}(2021)\citenamefont {Mishra},
  \citenamefont {Catarina}, \citenamefont {Wu}, \citenamefont {Ortiz},
  \citenamefont {Jacob}, \citenamefont {Eimre}, \citenamefont {Ma},
  \citenamefont {Pignedoli}, \citenamefont {Feng}, \citenamefont {Ruffieux},
  \citenamefont {Fernández-Rossier},\ and\ \citenamefont
  {Fasel}}]{mishra2021}%
  \BibitemOpen
  \bibfield  {author} {\bibinfo {author} {\bibfnamefont {S.}~\bibnamefont
  {Mishra}}, \bibinfo {author} {\bibfnamefont {G.}~\bibnamefont {Catarina}},
  \bibinfo {author} {\bibfnamefont {F.}~\bibnamefont {Wu}}, \bibinfo {author}
  {\bibfnamefont {R.}~\bibnamefont {Ortiz}}, \bibinfo {author} {\bibfnamefont
  {D.}~\bibnamefont {Jacob}}, \bibinfo {author} {\bibfnamefont
  {K.}~\bibnamefont {Eimre}}, \bibinfo {author} {\bibfnamefont
  {J.}~\bibnamefont {Ma}}, \bibinfo {author} {\bibfnamefont {C.~A.}\
  \bibnamefont {Pignedoli}}, \bibinfo {author} {\bibfnamefont {X.}~\bibnamefont
  {Feng}}, \bibinfo {author} {\bibfnamefont {P.}~\bibnamefont {Ruffieux}},
  \bibinfo {author} {\bibfnamefont {J.}~\bibnamefont {Fernández-Rossier}},\
  and\ \bibinfo {author} {\bibfnamefont {R.}~\bibnamefont {Fasel}},\ }\bibfield
   {title} {\bibinfo {title} {Observation of fractional edge excitations in
  nanographene spin chains},\ }\href
  {https://doi.org/10.1038/s41586-021-03842-3} {\bibfield  {journal} {\bibinfo
  {journal} {Nature}\ }\textbf {\bibinfo {volume} {598}},\ \bibinfo {pages}
  {287} (\bibinfo {year} {2021})}\BibitemShut {NoStop}%
\bibitem [{\citenamefont {Hagiwara}\ \emph
  {et~al.}(1990{\natexlab{b}})\citenamefont {Hagiwara}, \citenamefont
  {Katsumata}, \citenamefont {Affleck}, \citenamefont {Halperin},\ and\
  \citenamefont {Renard}}]{hagiwara1990}%
  \BibitemOpen
  \bibfield  {author} {\bibinfo {author} {\bibfnamefont {M.}~\bibnamefont
  {Hagiwara}}, \bibinfo {author} {\bibfnamefont {K.}~\bibnamefont {Katsumata}},
  \bibinfo {author} {\bibfnamefont {I.}~\bibnamefont {Affleck}}, \bibinfo
  {author} {\bibfnamefont {B.~I.}\ \bibnamefont {Halperin}},\ and\ \bibinfo
  {author} {\bibfnamefont {J.~P.}\ \bibnamefont {Renard}},\ }\bibfield  {title}
  {\bibinfo {title} {Observation of {{{\emph{S}}}} =1/2 degrees of freedom in
  an {{{\emph{S}}}} =1 linear-chain {{Heisenberg}} antiferromagnet},\ }\href
  {https://doi.org/10.1103/PhysRevLett.65.3181} {\bibfield  {journal} {\bibinfo
   {journal} {Phys. Rev. Lett.}\ }\textbf {\bibinfo {volume} {65}},\ \bibinfo
  {pages} {3181} (\bibinfo {year} {1990}{\natexlab{b}})}\BibitemShut {NoStop}%
\bibitem [{\citenamefont {Hewson}(1993)}]{Hewson_1993}%
  \BibitemOpen
  \bibfield  {author} {\bibinfo {author} {\bibfnamefont {A.}~\bibnamefont
  {Hewson}},\ }\href {https://doi.org/10.1017/CBO9780511470752} {\emph
  {\bibinfo {title} {The Kondo Problem to Heavy Fermions}}}\ (\bibinfo
  {publisher} {Cambridge University Press},\ \bibinfo {address} {Cambridge},\
  \bibinfo {year} {1993})\BibitemShut {NoStop}%
\bibitem [{\citenamefont {Trivini}\ \emph {et~al.}(2023)\citenamefont
  {Trivini}, \citenamefont {Ortuzar}, \citenamefont {Vaxevani}, \citenamefont
  {Li}, \citenamefont {Bergeret}, \citenamefont {Cazalilla},\ and\
  \citenamefont {Pascual}}]{trivini2023}%
  \BibitemOpen
  \bibfield  {author} {\bibinfo {author} {\bibfnamefont {S.}~\bibnamefont
  {Trivini}}, \bibinfo {author} {\bibfnamefont {J.}~\bibnamefont {Ortuzar}},
  \bibinfo {author} {\bibfnamefont {K.}~\bibnamefont {Vaxevani}}, \bibinfo
  {author} {\bibfnamefont {J.}~\bibnamefont {Li}}, \bibinfo {author}
  {\bibfnamefont {F.~S.}\ \bibnamefont {Bergeret}}, \bibinfo {author}
  {\bibfnamefont {M.~A.}\ \bibnamefont {Cazalilla}},\ and\ \bibinfo {author}
  {\bibfnamefont {J.~I.}\ \bibnamefont {Pascual}},\ }\bibfield  {title}
  {\bibinfo {title} {Cooper pair excitation mediated by a molecular quantum
  spin on a superconducting proximitized gold film},\ }\href
  {https://doi.org/10.1103/PhysRevLett.130.136004} {\bibfield  {journal}
  {\bibinfo  {journal} {Phys. Rev. Lett.}\ }\textbf {\bibinfo {volume} {130}},\
  \bibinfo {pages} {136004} (\bibinfo {year} {2023})}\BibitemShut {NoStop}%
\bibitem [{\citenamefont {Schneider}\ \emph {et~al.}(2024)\citenamefont
  {Schneider}, \citenamefont {von Bredow}, \citenamefont {Kim}, \citenamefont
  {Ton}, \citenamefont {Hänke}, \citenamefont {Wiebe},\ and\ \citenamefont
  {Wiesendanger}}]{schneider2024}%
  \BibitemOpen
  \bibfield  {author} {\bibinfo {author} {\bibfnamefont {L.}~\bibnamefont
  {Schneider}}, \bibinfo {author} {\bibfnamefont {C.}~\bibnamefont {von
  Bredow}}, \bibinfo {author} {\bibfnamefont {H.}~\bibnamefont {Kim}}, \bibinfo
  {author} {\bibfnamefont {K.~T.}\ \bibnamefont {Ton}}, \bibinfo {author}
  {\bibfnamefont {T.}~\bibnamefont {Hänke}}, \bibinfo {author} {\bibfnamefont
  {J.}~\bibnamefont {Wiebe}},\ and\ \bibinfo {author} {\bibfnamefont
  {R.}~\bibnamefont {Wiesendanger}},\ }\href
  {https://doi.org/10.48550/ARXIV.2402.08895} {\bibinfo {title}
  {High-resolution spectroscopy of proximity superconductivity in finite-size
  quantized surface states}} (\bibinfo {year} {2024}),\ \Eprint
  {https://arxiv.org/abs/2402.08895} {arXiv:2402.08895 [cond-mat.supr-con]}
  \BibitemShut {NoStop}%
\bibitem [{\citenamefont {Schneider}\ \emph {et~al.}(2023)\citenamefont
  {Schneider}, \citenamefont {Ton}, \citenamefont {Ioannidis}, \citenamefont
  {Neuhaus-Steinmetz}, \citenamefont {Posske}, \citenamefont {Wiesendanger},\
  and\ \citenamefont {Wiebe}}]{schneider2023}%
  \BibitemOpen
  \bibfield  {author} {\bibinfo {author} {\bibfnamefont {L.}~\bibnamefont
  {Schneider}}, \bibinfo {author} {\bibfnamefont {K.~T.}\ \bibnamefont {Ton}},
  \bibinfo {author} {\bibfnamefont {I.}~\bibnamefont {Ioannidis}}, \bibinfo
  {author} {\bibfnamefont {J.}~\bibnamefont {Neuhaus-Steinmetz}}, \bibinfo
  {author} {\bibfnamefont {T.}~\bibnamefont {Posske}}, \bibinfo {author}
  {\bibfnamefont {R.}~\bibnamefont {Wiesendanger}},\ and\ \bibinfo {author}
  {\bibfnamefont {J.}~\bibnamefont {Wiebe}},\ }\bibfield  {title} {\bibinfo
  {title} {Proximity superconductivity in atom-by-atom crafted quantum dots},\
  }\href {https://doi.org/10.1038/s41586-023-06312-0} {\bibfield  {journal}
  {\bibinfo  {journal} {Nature}\ }\textbf {\bibinfo {volume} {621}},\ \bibinfo
  {pages} {60} (\bibinfo {year} {2023})}\BibitemShut {NoStop}%
\bibitem [{\citenamefont {Vaxevani}\ \emph {et~al.}(2022)\citenamefont
  {Vaxevani}, \citenamefont {Li}, \citenamefont {Trivini}, \citenamefont
  {Ortuzar}, \citenamefont {Longo}, \citenamefont {Wang},\ and\ \citenamefont
  {Pascual}}]{vaxevani2022}%
  \BibitemOpen
  \bibfield  {author} {\bibinfo {author} {\bibfnamefont {K.}~\bibnamefont
  {Vaxevani}}, \bibinfo {author} {\bibfnamefont {J.}~\bibnamefont {Li}},
  \bibinfo {author} {\bibfnamefont {S.}~\bibnamefont {Trivini}}, \bibinfo
  {author} {\bibfnamefont {J.}~\bibnamefont {Ortuzar}}, \bibinfo {author}
  {\bibfnamefont {D.}~\bibnamefont {Longo}}, \bibinfo {author} {\bibfnamefont
  {D.}~\bibnamefont {Wang}},\ and\ \bibinfo {author} {\bibfnamefont {J.~I.}\
  \bibnamefont {Pascual}},\ }\bibfield  {title} {\bibinfo {title} {Extending
  the spin excitation lifetime of a magnetic molecule on a proximitized
  superconductor},\ }\href {https://doi.org/10.1021/acs.nanolett.2c00924}
  {\bibfield  {journal} {\bibinfo  {journal} {Nano Letters}\ }\textbf {\bibinfo
  {volume} {22}},\ \bibinfo {pages} {6075} (\bibinfo {year}
  {2022})}\BibitemShut {NoStop}%
\bibitem [{\citenamefont {Liu}\ \emph {et~al.}(2023)\citenamefont {Liu},
  \citenamefont {Pawlak}, \citenamefont {Wang}, \citenamefont {Chen},
  \citenamefont {D’Astolfo}, \citenamefont {Drechsel}, \citenamefont {Zhou},
  \citenamefont {Häner}, \citenamefont {Decurtins}, \citenamefont {Aschauer},
  \citenamefont {Liu}, \citenamefont {Wulfhekel},\ and\ \citenamefont
  {Meyer}}]{liu2023}%
  \BibitemOpen
  \bibfield  {author} {\bibinfo {author} {\bibfnamefont {J.-C.}\ \bibnamefont
  {Liu}}, \bibinfo {author} {\bibfnamefont {R.}~\bibnamefont {Pawlak}},
  \bibinfo {author} {\bibfnamefont {X.}~\bibnamefont {Wang}}, \bibinfo {author}
  {\bibfnamefont {H.}~\bibnamefont {Chen}}, \bibinfo {author} {\bibfnamefont
  {P.}~\bibnamefont {D’Astolfo}}, \bibinfo {author} {\bibfnamefont
  {C.}~\bibnamefont {Drechsel}}, \bibinfo {author} {\bibfnamefont
  {P.}~\bibnamefont {Zhou}}, \bibinfo {author} {\bibfnamefont {R.}~\bibnamefont
  {Häner}}, \bibinfo {author} {\bibfnamefont {S.}~\bibnamefont {Decurtins}},
  \bibinfo {author} {\bibfnamefont {U.}~\bibnamefont {Aschauer}}, \bibinfo
  {author} {\bibfnamefont {S.-X.}\ \bibnamefont {Liu}}, \bibinfo {author}
  {\bibfnamefont {W.}~\bibnamefont {Wulfhekel}},\ and\ \bibinfo {author}
  {\bibfnamefont {E.}~\bibnamefont {Meyer}},\ }\bibfield  {title} {\bibinfo
  {title} {Proximity-{{Induced Superconductivity}} in {{Atomically Precise
  Nanographene}} on {{Ag}}/{{Nb}}(110)},\ }\href
  {https://doi.org/10.1021/acsmaterialslett.2c00955} {\bibfield  {journal}
  {\bibinfo  {journal} {ACS Materials Lett.}\ }\textbf {\bibinfo {volume}
  {5}},\ \bibinfo {pages} {1083} (\bibinfo {year} {2023})}\BibitemShut
  {NoStop}%
\bibitem [{\citenamefont {Loss}\ and\ \citenamefont
  {DiVincenzo}(1998)}]{PhysRevA.57.120}%
  \BibitemOpen
  \bibfield  {author} {\bibinfo {author} {\bibfnamefont {D.}~\bibnamefont
  {Loss}}\ and\ \bibinfo {author} {\bibfnamefont {D.~P.}\ \bibnamefont
  {DiVincenzo}},\ }\bibfield  {title} {\bibinfo {title} {Quantum computation
  with quantum dots},\ }\href {https://doi.org/10.1103/PhysRevA.57.120}
  {\bibfield  {journal} {\bibinfo  {journal} {Phys. Rev. A}\ }\textbf {\bibinfo
  {volume} {57}},\ \bibinfo {pages} {120} (\bibinfo {year} {1998})}\BibitemShut
  {NoStop}%
\bibitem [{\citenamefont {Kloeffel}\ and\ \citenamefont
  {Loss}(2013)}]{Kloeffel2013}%
  \BibitemOpen
  \bibfield  {author} {\bibinfo {author} {\bibfnamefont {C.}~\bibnamefont
  {Kloeffel}}\ and\ \bibinfo {author} {\bibfnamefont {D.}~\bibnamefont
  {Loss}},\ }\bibfield  {title} {\bibinfo {title} {Prospects for spin-based
  quantum computing in quantum dots},\ }\href
  {https://doi.org/10.1146/annurev-conmatphys-030212-184248} {\bibfield
  {journal} {\bibinfo  {journal} {Annual Review of Condensed Matter Physics}\
  }\textbf {\bibinfo {volume} {4}},\ \bibinfo {pages} {51–81} (\bibinfo
  {year} {2013})}\BibitemShut {NoStop}%
\bibitem [{\citenamefont {Drexler}\ \emph {et~al.}(1994)\citenamefont
  {Drexler}, \citenamefont {Leonard}, \citenamefont {Hansen}, \citenamefont
  {Kotthaus},\ and\ \citenamefont {Petroff}}]{PhysRevLett.73.2252}%
  \BibitemOpen
  \bibfield  {author} {\bibinfo {author} {\bibfnamefont {H.}~\bibnamefont
  {Drexler}}, \bibinfo {author} {\bibfnamefont {D.}~\bibnamefont {Leonard}},
  \bibinfo {author} {\bibfnamefont {W.}~\bibnamefont {Hansen}}, \bibinfo
  {author} {\bibfnamefont {J.~P.}\ \bibnamefont {Kotthaus}},\ and\ \bibinfo
  {author} {\bibfnamefont {P.~M.}\ \bibnamefont {Petroff}},\ }\bibfield
  {title} {\bibinfo {title} {Spectroscopy of quantum levels in charge-tunable
  ingaas quantum dots},\ }\href {https://doi.org/10.1103/PhysRevLett.73.2252}
  {\bibfield  {journal} {\bibinfo  {journal} {Phys. Rev. Lett.}\ }\textbf
  {\bibinfo {volume} {73}},\ \bibinfo {pages} {2252} (\bibinfo {year}
  {1994})}\BibitemShut {NoStop}%
\bibitem [{\citenamefont {Nadj-Perge}\ \emph {et~al.}(2010)\citenamefont
  {Nadj-Perge}, \citenamefont {Frolov}, \citenamefont {Bakkers},\ and\
  \citenamefont {Kouwenhoven}}]{Nadj-Perge2010}%
  \BibitemOpen
  \bibfield  {author} {\bibinfo {author} {\bibfnamefont {S.}~\bibnamefont
  {Nadj-Perge}}, \bibinfo {author} {\bibfnamefont {S.~M.}\ \bibnamefont
  {Frolov}}, \bibinfo {author} {\bibfnamefont {E.~P. A.~M.}\ \bibnamefont
  {Bakkers}},\ and\ \bibinfo {author} {\bibfnamefont {L.~P.}\ \bibnamefont
  {Kouwenhoven}},\ }\bibfield  {title} {\bibinfo {title} {Spin--orbit qubit in
  a semiconductor nanowire},\ }\href {https://doi.org/10.1038/nature09682}
  {\bibfield  {journal} {\bibinfo  {journal} {Nature}\ }\textbf {\bibinfo
  {volume} {468}},\ \bibinfo {pages} {1084} (\bibinfo {year}
  {2010})}\BibitemShut {NoStop}%
\bibitem [{\citenamefont {van~den Berg}\ \emph {et~al.}(2013)\citenamefont
  {van~den Berg}, \citenamefont {Nadj-Perge}, \citenamefont {Pribiag},
  \citenamefont {Plissard}, \citenamefont {Bakkers}, \citenamefont {Frolov},\
  and\ \citenamefont {Kouwenhoven}}]{PhysRevLett.110.066806}%
  \BibitemOpen
  \bibfield  {author} {\bibinfo {author} {\bibfnamefont {J.~W.~G.}\
  \bibnamefont {van~den Berg}}, \bibinfo {author} {\bibfnamefont
  {S.}~\bibnamefont {Nadj-Perge}}, \bibinfo {author} {\bibfnamefont {V.~S.}\
  \bibnamefont {Pribiag}}, \bibinfo {author} {\bibfnamefont {S.~R.}\
  \bibnamefont {Plissard}}, \bibinfo {author} {\bibfnamefont {E.~P. A.~M.}\
  \bibnamefont {Bakkers}}, \bibinfo {author} {\bibfnamefont {S.~M.}\
  \bibnamefont {Frolov}},\ and\ \bibinfo {author} {\bibfnamefont {L.~P.}\
  \bibnamefont {Kouwenhoven}},\ }\bibfield  {title} {\bibinfo {title} {Fast
  spin-orbit qubit in an indium antimonide nanowire},\ }\href
  {https://doi.org/10.1103/PhysRevLett.110.066806} {\bibfield  {journal}
  {\bibinfo  {journal} {Phys. Rev. Lett.}\ }\textbf {\bibinfo {volume} {110}},\
  \bibinfo {pages} {066806} (\bibinfo {year} {2013})}\BibitemShut {NoStop}%
\bibitem [{\citenamefont {Scheibner}\ \emph {et~al.}(2008)\citenamefont
  {Scheibner}, \citenamefont {Yakes}, \citenamefont {Bracker}, \citenamefont
  {Ponomarev}, \citenamefont {Doty}, \citenamefont {Hellberg}, \citenamefont
  {Whitman}, \citenamefont {Reinecke},\ and\ \citenamefont
  {Gammon}}]{Scheibner2008}%
  \BibitemOpen
  \bibfield  {author} {\bibinfo {author} {\bibfnamefont {M.}~\bibnamefont
  {Scheibner}}, \bibinfo {author} {\bibfnamefont {M.}~\bibnamefont {Yakes}},
  \bibinfo {author} {\bibfnamefont {A.~S.}\ \bibnamefont {Bracker}}, \bibinfo
  {author} {\bibfnamefont {I.~V.}\ \bibnamefont {Ponomarev}}, \bibinfo {author}
  {\bibfnamefont {M.~F.}\ \bibnamefont {Doty}}, \bibinfo {author}
  {\bibfnamefont {C.~S.}\ \bibnamefont {Hellberg}}, \bibinfo {author}
  {\bibfnamefont {L.~J.}\ \bibnamefont {Whitman}}, \bibinfo {author}
  {\bibfnamefont {T.~L.}\ \bibnamefont {Reinecke}},\ and\ \bibinfo {author}
  {\bibfnamefont {D.}~\bibnamefont {Gammon}},\ }\bibfield  {title} {\bibinfo
  {title} {Optically mapping the electronic structure of coupled quantum
  dots},\ }\href {https://doi.org/10.1038/nphys882} {\bibfield  {journal}
  {\bibinfo  {journal} {Nature Physics}\ }\textbf {\bibinfo {volume} {4}},\
  \bibinfo {pages} {291} (\bibinfo {year} {2008})}\BibitemShut {NoStop}%
\bibitem [{\citenamefont {{Gaita-Ari{\~n}o}}\ \emph {et~al.}(2019)\citenamefont
  {{Gaita-Ari{\~n}o}}, \citenamefont {Luis}, \citenamefont {Hill},\ and\
  \citenamefont {Coronado}}]{Gaita-Arino2019}%
  \BibitemOpen
  \bibfield  {author} {\bibinfo {author} {\bibfnamefont {A.}~\bibnamefont
  {{Gaita-Ari{\~n}o}}}, \bibinfo {author} {\bibfnamefont {F.}~\bibnamefont
  {Luis}}, \bibinfo {author} {\bibfnamefont {S.}~\bibnamefont {Hill}},\ and\
  \bibinfo {author} {\bibfnamefont {E.}~\bibnamefont {Coronado}},\ }\bibfield
  {title} {\bibinfo {title} {Molecular spins for quantum computation},\ }\href
  {https://doi.org/10.1038/s41557-019-0232-y} {\bibfield  {journal} {\bibinfo
  {journal} {Nature Chemistry}\ }\textbf {\bibinfo {volume} {11}},\ \bibinfo
  {pages} {301} (\bibinfo {year} {2019})}\BibitemShut {NoStop}%
\bibitem [{\citenamefont {Bonesteel}\ \emph {et~al.}(2001)\citenamefont
  {Bonesteel}, \citenamefont {Stepanenko},\ and\ \citenamefont
  {DiVincenzo}}]{PhysRevLett.87.207901}%
  \BibitemOpen
  \bibfield  {author} {\bibinfo {author} {\bibfnamefont {N.~E.}\ \bibnamefont
  {Bonesteel}}, \bibinfo {author} {\bibfnamefont {D.}~\bibnamefont
  {Stepanenko}},\ and\ \bibinfo {author} {\bibfnamefont {D.~P.}\ \bibnamefont
  {DiVincenzo}},\ }\bibfield  {title} {\bibinfo {title} {Anisotropic spin
  exchange in pulsed quantum gates},\ }\href
  {https://doi.org/10.1103/PhysRevLett.87.207901} {\bibfield  {journal}
  {\bibinfo  {journal} {Phys. Rev. Lett.}\ }\textbf {\bibinfo {volume} {87}},\
  \bibinfo {pages} {207901} (\bibinfo {year} {2001})}\BibitemShut {NoStop}%
\bibitem [{\citenamefont {Burkard}\ and\ \citenamefont
  {Loss}(2002)}]{PhysRevLett.88.047903}%
  \BibitemOpen
  \bibfield  {author} {\bibinfo {author} {\bibfnamefont {G.}~\bibnamefont
  {Burkard}}\ and\ \bibinfo {author} {\bibfnamefont {D.}~\bibnamefont {Loss}},\
  }\bibfield  {title} {\bibinfo {title} {Cancellation of spin-orbit effects in
  quantum gates based on the exchange coupling in quantum dots},\ }\href
  {https://doi.org/10.1103/PhysRevLett.88.047903} {\bibfield  {journal}
  {\bibinfo  {journal} {Phys. Rev. Lett.}\ }\textbf {\bibinfo {volume} {88}},\
  \bibinfo {pages} {047903} (\bibinfo {year} {2002})}\BibitemShut {NoStop}%
\bibitem [{\citenamefont {Coish}\ and\ \citenamefont
  {Loss}(2004)}]{PhysRevB.70.195340}%
  \BibitemOpen
  \bibfield  {author} {\bibinfo {author} {\bibfnamefont {W.~A.}\ \bibnamefont
  {Coish}}\ and\ \bibinfo {author} {\bibfnamefont {D.}~\bibnamefont {Loss}},\
  }\bibfield  {title} {\bibinfo {title} {Hyperfine interaction in a quantum
  dot: Non-markovian electron spin dynamics},\ }\href
  {https://doi.org/10.1103/PhysRevB.70.195340} {\bibfield  {journal} {\bibinfo
  {journal} {Phys. Rev. B}\ }\textbf {\bibinfo {volume} {70}},\ \bibinfo
  {pages} {195340} (\bibinfo {year} {2004})}\BibitemShut {NoStop}%
\bibitem [{\citenamefont {Fischer}\ \emph {et~al.}(2008)\citenamefont
  {Fischer}, \citenamefont {Coish}, \citenamefont {Bulaev},\ and\ \citenamefont
  {Loss}}]{PhysRevB.78.155329}%
  \BibitemOpen
  \bibfield  {author} {\bibinfo {author} {\bibfnamefont {J.}~\bibnamefont
  {Fischer}}, \bibinfo {author} {\bibfnamefont {W.~A.}\ \bibnamefont {Coish}},
  \bibinfo {author} {\bibfnamefont {D.~V.}\ \bibnamefont {Bulaev}},\ and\
  \bibinfo {author} {\bibfnamefont {D.}~\bibnamefont {Loss}},\ }\bibfield
  {title} {\bibinfo {title} {Spin decoherence of a heavy hole coupled to
  nuclear spins in a quantum dot},\ }\href
  {https://doi.org/10.1103/PhysRevB.78.155329} {\bibfield  {journal} {\bibinfo
  {journal} {Phys. Rev. B}\ }\textbf {\bibinfo {volume} {78}},\ \bibinfo
  {pages} {155329} (\bibinfo {year} {2008})}\BibitemShut {NoStop}%
\bibitem [{\citenamefont {Danon}\ \emph {et~al.}(2021)\citenamefont {Danon},
  \citenamefont {Chatterjee}, \citenamefont {Gyenis},\ and\ \citenamefont
  {Kuemmeth}}]{Danon2021}%
  \BibitemOpen
  \bibfield  {author} {\bibinfo {author} {\bibfnamefont {J.}~\bibnamefont
  {Danon}}, \bibinfo {author} {\bibfnamefont {A.}~\bibnamefont {Chatterjee}},
  \bibinfo {author} {\bibfnamefont {A.}~\bibnamefont {Gyenis}},\ and\ \bibinfo
  {author} {\bibfnamefont {F.}~\bibnamefont {Kuemmeth}},\ }\bibfield  {title}
  {\bibinfo {title} {Protected solid-state qubits},\ }\href
  {https://doi.org/10.1063/5.0073945} {\bibfield  {journal} {\bibinfo
  {journal} {Applied Physics Letters}\ }\textbf {\bibinfo {volume} {119}},\
  \bibinfo {pages} {260502} (\bibinfo {year} {2021})}\BibitemShut {NoStop}%
\bibitem [{\citenamefont {Yao}\ \emph {et~al.}(2014{\natexlab{a}})\citenamefont
  {Yao}, \citenamefont {Moca}, \citenamefont {Weymann}, \citenamefont {Sau},
  \citenamefont {Lukin}, \citenamefont {Demler},\ and\ \citenamefont
  {Zar\'and}}]{yao2014}%
  \BibitemOpen
  \bibfield  {author} {\bibinfo {author} {\bibfnamefont {N.~Y.}\ \bibnamefont
  {Yao}}, \bibinfo {author} {\bibfnamefont {C.~P.}\ \bibnamefont {Moca}},
  \bibinfo {author} {\bibfnamefont {I.}~\bibnamefont {Weymann}}, \bibinfo
  {author} {\bibfnamefont {J.~D.}\ \bibnamefont {Sau}}, \bibinfo {author}
  {\bibfnamefont {M.~D.}\ \bibnamefont {Lukin}}, \bibinfo {author}
  {\bibfnamefont {E.~A.}\ \bibnamefont {Demler}},\ and\ \bibinfo {author}
  {\bibfnamefont {G.}~\bibnamefont {Zar\'and}},\ }\bibfield  {title} {\bibinfo
  {title} {Phase diagram and excitations of a shiba molecule},\ }\href
  {https://doi.org/10.1103/PhysRevB.90.241108} {\bibfield  {journal} {\bibinfo
  {journal} {Phys. Rev. B}\ }\textbf {\bibinfo {volume} {90}},\ \bibinfo
  {pages} {241108} (\bibinfo {year} {2014}{\natexlab{a}})}\BibitemShut
  {NoStop}%
\bibitem [{\citenamefont {Saleem}\ \emph {et~al.}(2024)\citenamefont {Saleem},
  \citenamefont {Steenbock}, \citenamefont {Alhadi}, \citenamefont {Pasek},
  \citenamefont {Bester},\ and\ \citenamefont
  {Potasz}}]{Saleem_et_al_NanoLett_2024}%
  \BibitemOpen
  \bibfield  {author} {\bibinfo {author} {\bibfnamefont {Y.}~\bibnamefont
  {Saleem}}, \bibinfo {author} {\bibfnamefont {T.}~\bibnamefont {Steenbock}},
  \bibinfo {author} {\bibfnamefont {E.~R.~J.}\ \bibnamefont {Alhadi}}, \bibinfo
  {author} {\bibfnamefont {W.}~\bibnamefont {Pasek}}, \bibinfo {author}
  {\bibfnamefont {G.}~\bibnamefont {Bester}},\ and\ \bibinfo {author}
  {\bibfnamefont {P.}~\bibnamefont {Potasz}},\ }\bibfield  {title} {\bibinfo
  {title} {Superexchange mechanism in coupled triangulenes forming spin-1
  chains},\ }\href {https://doi.org/10.1021/acs.nanolett.4c01604} {\bibfield
  {journal} {\bibinfo  {journal} {Nano Letters}\ }\textbf {\bibinfo {volume}
  {24}},\ \bibinfo {pages} {7417} (\bibinfo {year} {2024})}\BibitemShut
  {NoStop}%
\bibitem [{Note1()}]{Note1}%
  \BibitemOpen
  \bibinfo {note} {$J_\protect \text {RKKY}\sim \Delta e^{-2L/\xi _\protect
  \text {SC}} \protect \frac {\cos ^2(k_FL)}{2(k_FL)^2}$~\cite
  {PhysRevLett.113.087202RKKY}. Taking some realistic values, corresponding to
  a Nb sample, such as $\Delta \sim 2.5$~meV, $\xi _{SC}\simeq 40$~nm, and
  $k_F\sim 13.7 \protect \text {nm}^{-1} $, we have $J_\protect \text
  {eff}/J_\protect \text {RKKY}\gg 1$ for the experimentally observed
  triangulene chains (5 to 15 units)}\BibitemShut {NoStop}%
\bibitem [{\citenamefont {Franke}\ \emph {et~al.}(2011)\citenamefont {Franke},
  \citenamefont {Schulze},\ and\ \citenamefont {Pascual}}]{franke2011}%
  \BibitemOpen
  \bibfield  {author} {\bibinfo {author} {\bibfnamefont {K.~J.}\ \bibnamefont
  {Franke}}, \bibinfo {author} {\bibfnamefont {G.}~\bibnamefont {Schulze}},\
  and\ \bibinfo {author} {\bibfnamefont {J.~I.}\ \bibnamefont {Pascual}},\
  }\bibfield  {title} {\bibinfo {title} {Competition of superconducting
  phenomena and kondo screening at the nanoscale},\ }\href
  {https://doi.org/10.1126/science.1202204} {\bibfield  {journal} {\bibinfo
  {journal} {Science}\ }\textbf {\bibinfo {volume} {332}},\ \bibinfo {pages}
  {940} (\bibinfo {year} {2011})}\BibitemShut {NoStop}%
\bibitem [{\citenamefont {Karan}\ \emph {et~al.}(2024)\citenamefont {Karan},
  \citenamefont {Huang}, \citenamefont {Ivanovic}, \citenamefont {Padurariu},
  \citenamefont {Kubala}, \citenamefont {Kern}, \citenamefont {Ankerhold},\
  and\ \citenamefont {Ast}}]{karan2024}%
  \BibitemOpen
  \bibfield  {author} {\bibinfo {author} {\bibfnamefont {S.}~\bibnamefont
  {Karan}}, \bibinfo {author} {\bibfnamefont {H.}~\bibnamefont {Huang}},
  \bibinfo {author} {\bibfnamefont {A.}~\bibnamefont {Ivanovic}}, \bibinfo
  {author} {\bibfnamefont {C.}~\bibnamefont {Padurariu}}, \bibinfo {author}
  {\bibfnamefont {B.}~\bibnamefont {Kubala}}, \bibinfo {author} {\bibfnamefont
  {K.}~\bibnamefont {Kern}}, \bibinfo {author} {\bibfnamefont {J.}~\bibnamefont
  {Ankerhold}},\ and\ \bibinfo {author} {\bibfnamefont {C.~R.}\ \bibnamefont
  {Ast}},\ }\bibfield  {title} {\bibinfo {title} {Tracking a spin-polarized
  superconducting bound state across a quantum phase transition},\ }\href
  {https://doi.org/10.1038/s41467-024-44708-2} {\bibfield  {journal} {\bibinfo
  {journal} {Nat Commun}\ }\textbf {\bibinfo {volume} {15}},\ \bibinfo {pages}
  {459} (\bibinfo {year} {2024})}\BibitemShut {NoStop}%
\bibitem [{\citenamefont {van Mullekom}\ \emph {et~al.}(2024)\citenamefont {van
  Mullekom}, \citenamefont {Verlhac}, \citenamefont {van Weerdenburg},
  \citenamefont {Osterhage}, \citenamefont {Steinbrecher}, \citenamefont
  {Franke},\ and\ \citenamefont {Khajetoorians}}]{Niels2024}%
  \BibitemOpen
  \bibfield  {author} {\bibinfo {author} {\bibfnamefont {N.~P.~E.}\
  \bibnamefont {van Mullekom}}, \bibinfo {author} {\bibfnamefont
  {B.}~\bibnamefont {Verlhac}}, \bibinfo {author} {\bibfnamefont {W.~M.~J.}\
  \bibnamefont {van Weerdenburg}}, \bibinfo {author} {\bibfnamefont
  {H.}~\bibnamefont {Osterhage}}, \bibinfo {author} {\bibfnamefont
  {M.}~\bibnamefont {Steinbrecher}}, \bibinfo {author} {\bibfnamefont {K.~J.}\
  \bibnamefont {Franke}},\ and\ \bibinfo {author} {\bibfnamefont {A.~A.}\
  \bibnamefont {Khajetoorians}},\ }\bibfield  {title} {\bibinfo {title}
  {Quantifying the quantum nature of high-spin ysr excitations in transverse
  magnetic field},\ }\href {https://doi.org/10.1126/sciadv.adq0965} {\bibfield
  {journal} {\bibinfo  {journal} {Science Advances}\ }\textbf {\bibinfo
  {volume} {10}},\ \bibinfo {pages} {eadq0965} (\bibinfo {year}
  {2024})}\BibitemShut {NoStop}%
\bibitem [{Note2()}]{Note2}%
  \BibitemOpen
  \bibinfo {note} {This regime may be accessible in open ring TSCs for which
  the edge distance is small enough for the Kondo exchange of both terminal
  units to be tuned using a single STM tip.}\BibitemShut {Stop}%
\bibitem [{\citenamefont {S\'anchez}\ \emph
  {et~al.}(2014{\natexlab{a}})\citenamefont {S\'anchez}, \citenamefont
  {Granger}, \citenamefont {Gaudreau}, \citenamefont {Kam}, \citenamefont
  {Pioro-Ladri\`ere}, \citenamefont {Studenikin}, \citenamefont {Zawadzki},
  \citenamefont {Sachrajda},\ and\ \citenamefont {Platero}}]{Sanchez2014}%
  \BibitemOpen
  \bibfield  {author} {\bibinfo {author} {\bibfnamefont {R.}~\bibnamefont
  {S\'anchez}}, \bibinfo {author} {\bibfnamefont {G.}~\bibnamefont {Granger}},
  \bibinfo {author} {\bibfnamefont {L.}~\bibnamefont {Gaudreau}}, \bibinfo
  {author} {\bibfnamefont {A.}~\bibnamefont {Kam}}, \bibinfo {author}
  {\bibfnamefont {M.}~\bibnamefont {Pioro-Ladri\`ere}}, \bibinfo {author}
  {\bibfnamefont {S.~A.}\ \bibnamefont {Studenikin}}, \bibinfo {author}
  {\bibfnamefont {P.}~\bibnamefont {Zawadzki}}, \bibinfo {author}
  {\bibfnamefont {A.~S.}\ \bibnamefont {Sachrajda}},\ and\ \bibinfo {author}
  {\bibfnamefont {G.}~\bibnamefont {Platero}},\ }\bibfield  {title} {\bibinfo
  {title} {Long-range spin transfer in triple quantum dots},\ }\href
  {https://doi.org/10.1103/PhysRevLett.112.176803} {\bibfield  {journal}
  {\bibinfo  {journal} {Phys. Rev. Lett.}\ }\textbf {\bibinfo {volume} {112}},\
  \bibinfo {pages} {176803} (\bibinfo {year} {2014}{\natexlab{a}})}\BibitemShut
  {NoStop}%
\bibitem [{\citenamefont {Baart}\ \emph {et~al.}(2017)\citenamefont {Baart},
  \citenamefont {Fujita}, \citenamefont {Reichl}, \citenamefont {Wegscheider},\
  and\ \citenamefont {Vandersypen}}]{baart2017}%
  \BibitemOpen
  \bibfield  {author} {\bibinfo {author} {\bibfnamefont {T.~A.}\ \bibnamefont
  {Baart}}, \bibinfo {author} {\bibfnamefont {T.}~\bibnamefont {Fujita}},
  \bibinfo {author} {\bibfnamefont {C.}~\bibnamefont {Reichl}}, \bibinfo
  {author} {\bibfnamefont {W.}~\bibnamefont {Wegscheider}},\ and\ \bibinfo
  {author} {\bibfnamefont {L.~M.~K.}\ \bibnamefont {Vandersypen}},\ }\bibfield
  {title} {\bibinfo {title} {Coherent spin-exchange via a quantum mediator},\
  }\href {https://doi.org/10.1038/nnano.2016.188} {\bibfield  {journal}
  {\bibinfo  {journal} {Nature Nanotech}\ }\textbf {\bibinfo {volume} {12}},\
  \bibinfo {pages} {26} (\bibinfo {year} {2017})}\BibitemShut {NoStop}%
\bibitem [{\citenamefont {Yu}(1965)}]{yu1965}%
  \BibitemOpen
  \bibfield  {author} {\bibinfo {author} {\bibfnamefont {L.}~\bibnamefont
  {Yu}},\ }\bibfield  {title} {\bibinfo {title} {{{Bound State in
  Superconductors with Paramagnetic Impurities}}},\ }\href
  {https://doi.org/10.7498/aps.21.75} {\bibfield  {journal} {\bibinfo
  {journal} {Acta Physica Sinica}\ }\textbf {\bibinfo {volume} {21}},\ \bibinfo
  {pages} {75} (\bibinfo {year} {1965})}\BibitemShut {NoStop}%
\bibitem [{\citenamefont {Shiba}(1968)}]{shiba1968}%
  \BibitemOpen
  \bibfield  {author} {\bibinfo {author} {\bibfnamefont {H.}~\bibnamefont
  {Shiba}},\ }\bibfield  {title} {\bibinfo {title} {Classical {{Spins}} in
  {{Superconductors}}},\ }\href {https://doi.org/10.1143/PTP.40.435} {\bibfield
   {journal} {\bibinfo  {journal} {Prog. Theor. Phys.}\ }\textbf {\bibinfo
  {volume} {40}},\ \bibinfo {pages} {435} (\bibinfo {year} {1968})}\BibitemShut
  {NoStop}%
\bibitem [{\citenamefont {Rusinov}(1969)}]{rusinov1969}%
  \BibitemOpen
  \bibfield  {author} {\bibinfo {author} {\bibfnamefont {A.~I.}\ \bibnamefont
  {Rusinov}},\ }\bibfield  {title} {\bibinfo {title} {On the {{Theory}} of
  {{Gapless Superconductivity}} in {{Alloys Containing Paramagnetic
  Impurities}}},\ }\href@noop {} {\bibfield  {journal} {\bibinfo  {journal}
  {Sov. J. Exp. Theor. Phys.}\ }\textbf {\bibinfo {volume} {29}},\ \bibinfo
  {pages} {1101} (\bibinfo {year} {1969})}\BibitemShut {NoStop}%
\bibitem [{\citenamefont {von Oppen}\ and\ \citenamefont
  {Franke}(2021)}]{vonoppen1}%
  \BibitemOpen
  \bibfield  {author} {\bibinfo {author} {\bibfnamefont {F.}~\bibnamefont {von
  Oppen}}\ and\ \bibinfo {author} {\bibfnamefont {K.~J.}\ \bibnamefont
  {Franke}},\ }\bibfield  {title} {\bibinfo {title} {Yu-shiba-rusinov states in
  real metals},\ }\href {https://doi.org/10.1103/PhysRevB.103.205424}
  {\bibfield  {journal} {\bibinfo  {journal} {Phys. Rev. B}\ }\textbf {\bibinfo
  {volume} {103}},\ \bibinfo {pages} {205424} (\bibinfo {year}
  {2021})}\BibitemShut {NoStop}%
\bibitem [{\citenamefont {Schmid}\ \emph {et~al.}(2022)\citenamefont {Schmid},
  \citenamefont {Steiner}, \citenamefont {Franke},\ and\ \citenamefont {{von
  Oppen}}}]{schmid2022}%
  \BibitemOpen
  \bibfield  {author} {\bibinfo {author} {\bibfnamefont {H.}~\bibnamefont
  {Schmid}}, \bibinfo {author} {\bibfnamefont {J.~F.}\ \bibnamefont {Steiner}},
  \bibinfo {author} {\bibfnamefont {K.~J.}\ \bibnamefont {Franke}},\ and\
  \bibinfo {author} {\bibfnamefont {F.}~\bibnamefont {{von Oppen}}},\
  }\bibfield  {title} {\bibinfo {title} {Quantum yu-shiba-rusinov dimers},\
  }\href {https://doi.org/10.1103/PhysRevB.105.235406} {\bibfield  {journal}
  {\bibinfo  {journal} {Phys. Rev. B}\ }\textbf {\bibinfo {volume} {105}},\
  \bibinfo {pages} {235406} (\bibinfo {year} {2022})}\BibitemShut {NoStop}%
\bibitem [{Note3()}]{Note3}%
  \BibitemOpen
  \bibinfo {note} {An antenna emitting a microwave field could be used to
  control the qubit, similar to electron-spin resonance STM or atomic-force
  microscope (AFM) measurements in Refs.~\cite {Baumann2015,Sellies2023}.
  However, new detection mechanisms would be required, as the singlet nature of
  the qubit prevents the readout via magnetoresistance changes~\cite
  {Baumann2015}. Charge state variations could still be used~\cite
  {Sellies2023}}\BibitemShut {NoStop}%
\bibitem [{\citenamefont {Deng}\ and\ \citenamefont
  {Barnes}(2020)}]{Deng_PhysRevB.102.035427}%
  \BibitemOpen
  \bibfield  {author} {\bibinfo {author} {\bibfnamefont {K.}~\bibnamefont
  {Deng}}\ and\ \bibinfo {author} {\bibfnamefont {E.}~\bibnamefont {Barnes}},\
  }\bibfield  {title} {\bibinfo {title} {Interplay of exchange and
  superexchange in triple quantum dots},\ }\href
  {https://doi.org/10.1103/PhysRevB.102.035427} {\bibfield  {journal} {\bibinfo
   {journal} {Phys. Rev. B}\ }\textbf {\bibinfo {volume} {102}},\ \bibinfo
  {pages} {035427} (\bibinfo {year} {2020})}\BibitemShut {NoStop}%
\bibitem [{\citenamefont {S\'anchez}\ \emph
  {et~al.}(2014{\natexlab{b}})\citenamefont {S\'anchez}, \citenamefont
  {Gallego-Marcos},\ and\ \citenamefont {Platero}}]{Sanchez2014b}%
  \BibitemOpen
  \bibfield  {author} {\bibinfo {author} {\bibfnamefont {R.}~\bibnamefont
  {S\'anchez}}, \bibinfo {author} {\bibfnamefont {F.}~\bibnamefont
  {Gallego-Marcos}},\ and\ \bibinfo {author} {\bibfnamefont {G.}~\bibnamefont
  {Platero}},\ }\bibfield  {title} {\bibinfo {title} {Superexchange blockade in
  triple quantum dots},\ }\href {https://doi.org/10.1103/PhysRevB.89.161402}
  {\bibfield  {journal} {\bibinfo  {journal} {Phys. Rev. B}\ }\textbf {\bibinfo
  {volume} {89}},\ \bibinfo {pages} {161402} (\bibinfo {year}
  {2014}{\natexlab{b}})}\BibitemShut {NoStop}%
\bibitem [{\citenamefont {Steffensen}\ and\ \citenamefont
  {Yeyati}(2025)}]{steffensen2024ysrbondqubitdouble_bond_q}%
  \BibitemOpen
  \bibfield  {author} {\bibinfo {author} {\bibfnamefont {G.~O.}\ \bibnamefont
  {Steffensen}}\ and\ \bibinfo {author} {\bibfnamefont {A.~L.}\ \bibnamefont
  {Yeyati}},\ }\bibfield  {title} {\bibinfo {title} {Yu-shiba-rusinov--bond
  qubit in a double quantum dot with circuit-qed operation},\ }\href
  {https://doi.org/10.1103/PRXQuantum.6.020329} {\bibfield  {journal} {\bibinfo
   {journal} {PRX Quantum}\ }\textbf {\bibinfo {volume} {6}},\ \bibinfo {pages}
  {020329} (\bibinfo {year} {2025})}\BibitemShut {NoStop}%
\bibitem [{Note4()}]{Note4}%
  \BibitemOpen
  \bibinfo {note} {Our NRG calculations of double quantum-dot system coupled by
  tunneling (rather than super-exchange) confirm the results of Ref.~\protect
  \rev@citealp {steffensen2024ysrbondqubitdouble_bond_q}, which were obtained
  in the infinite gap limit and yield a low-lying spectrum where the odd
  fermion-parity doublets lie between the even fermion-parity singlets~\cite
  {unpub2025}.}\BibitemShut {Stop}%
\bibitem [{\citenamefont {Pan}\ \emph {et~al.}()\citenamefont {Pan},
  \citenamefont {Zhou}, \citenamefont {Yuan}, \citenamefont {Nie},
  \citenamefont {Wei}, \citenamefont {Zhang}, \citenamefont {Li}, \citenamefont
  {Liu}, \citenamefont {Jiang}, \citenamefont {Catelani}, \citenamefont {Hu},
  \citenamefont {Yan},\ and\ \citenamefont {Yu}}]{pan_engineering_2022}%
  \BibitemOpen
  \bibfield  {author} {\bibinfo {author} {\bibfnamefont {X.}~\bibnamefont
  {Pan}}, \bibinfo {author} {\bibfnamefont {Y.}~\bibnamefont {Zhou}}, \bibinfo
  {author} {\bibfnamefont {H.}~\bibnamefont {Yuan}}, \bibinfo {author}
  {\bibfnamefont {L.}~\bibnamefont {Nie}}, \bibinfo {author} {\bibfnamefont
  {W.}~\bibnamefont {Wei}}, \bibinfo {author} {\bibfnamefont {L.}~\bibnamefont
  {Zhang}}, \bibinfo {author} {\bibfnamefont {J.}~\bibnamefont {Li}}, \bibinfo
  {author} {\bibfnamefont {S.}~\bibnamefont {Liu}}, \bibinfo {author}
  {\bibfnamefont {Z.~H.}\ \bibnamefont {Jiang}}, \bibinfo {author}
  {\bibfnamefont {G.}~\bibnamefont {Catelani}}, \bibinfo {author}
  {\bibfnamefont {L.}~\bibnamefont {Hu}}, \bibinfo {author} {\bibfnamefont
  {F.}~\bibnamefont {Yan}},\ and\ \bibinfo {author} {\bibfnamefont
  {D.}~\bibnamefont {Yu}},\ }\bibfield  {title} {\bibinfo {title} {Engineering
  superconducting qubits to reduce quasiparticles and charge noise},\ }\href
  {https://doi.org/10.1038/s41467-022-34727-2} {\ \textbf {\bibinfo {volume}
  {13}},\ \bibinfo {pages} {7196}}\BibitemShut {NoStop}%
\bibitem [{\citenamefont {Riwar}\ \emph {et~al.}(2016)\citenamefont {Riwar},
  \citenamefont {Hosseinkhani}, \citenamefont {Burkhart}, \citenamefont {Gao},
  \citenamefont {Schoelkopf}, \citenamefont {Glazman},\ and\ \citenamefont
  {Catelani}}]{PhysRevB.94.104516}%
  \BibitemOpen
  \bibfield  {author} {\bibinfo {author} {\bibfnamefont {R.-P.}\ \bibnamefont
  {Riwar}}, \bibinfo {author} {\bibfnamefont {A.}~\bibnamefont {Hosseinkhani}},
  \bibinfo {author} {\bibfnamefont {L.~D.}\ \bibnamefont {Burkhart}}, \bibinfo
  {author} {\bibfnamefont {Y.~Y.}\ \bibnamefont {Gao}}, \bibinfo {author}
  {\bibfnamefont {R.~J.}\ \bibnamefont {Schoelkopf}}, \bibinfo {author}
  {\bibfnamefont {L.~I.}\ \bibnamefont {Glazman}},\ and\ \bibinfo {author}
  {\bibfnamefont {G.}~\bibnamefont {Catelani}},\ }\bibfield  {title} {\bibinfo
  {title} {Normal-metal quasiparticle traps for superconducting qubits},\
  }\href {https://doi.org/10.1103/PhysRevB.94.104516} {\bibfield  {journal}
  {\bibinfo  {journal} {Phys. Rev. B}\ }\textbf {\bibinfo {volume} {94}},\
  \bibinfo {pages} {104516} (\bibinfo {year} {2016})}\BibitemShut {NoStop}%
\bibitem [{\citenamefont {Olivares}\ \emph {et~al.}(2014)\citenamefont
  {Olivares}, \citenamefont {Yeyati}, \citenamefont {Bretheau}, \citenamefont
  {Girit}, \citenamefont {Pothier},\ and\ \citenamefont
  {Urbina}}]{PhysRevB.89.104504}%
  \BibitemOpen
  \bibfield  {author} {\bibinfo {author} {\bibfnamefont {D.~G.}\ \bibnamefont
  {Olivares}}, \bibinfo {author} {\bibfnamefont {A.~L.}\ \bibnamefont
  {Yeyati}}, \bibinfo {author} {\bibfnamefont {L.}~\bibnamefont {Bretheau}},
  \bibinfo {author} {\bibfnamefont {{\c{C}}.~{\"{O}}.}\ \bibnamefont {Girit}},
  \bibinfo {author} {\bibfnamefont {H.}~\bibnamefont {Pothier}},\ and\ \bibinfo
  {author} {\bibfnamefont {C.}~\bibnamefont {Urbina}},\ }\bibfield  {title}
  {\bibinfo {title} {Dynamics of quasiparticle trapping in andreev levels},\
  }\href {https://doi.org/10.1103/PhysRevB.89.104504} {\bibfield  {journal}
  {\bibinfo  {journal} {Phys. Rev. B}\ }\textbf {\bibinfo {volume} {89}},\
  \bibinfo {pages} {104504} (\bibinfo {year} {2014})}\BibitemShut {NoStop}%
\bibitem [{\citenamefont {Ithier}\ \emph {et~al.}(2005)\citenamefont {Ithier},
  \citenamefont {Collin}, \citenamefont {Joyez}, \citenamefont {Meeson},
  \citenamefont {Vion}, \citenamefont {Esteve}, \citenamefont {Chiarello},
  \citenamefont {Shnirman}, \citenamefont {Makhlin}, \citenamefont {Schriefl},\
  and\ \citenamefont {Sch\"on}}]{PhysRevB.72.134519}%
  \BibitemOpen
  \bibfield  {author} {\bibinfo {author} {\bibfnamefont {G.}~\bibnamefont
  {Ithier}}, \bibinfo {author} {\bibfnamefont {E.}~\bibnamefont {Collin}},
  \bibinfo {author} {\bibfnamefont {P.}~\bibnamefont {Joyez}}, \bibinfo
  {author} {\bibfnamefont {P.~J.}\ \bibnamefont {Meeson}}, \bibinfo {author}
  {\bibfnamefont {D.}~\bibnamefont {Vion}}, \bibinfo {author} {\bibfnamefont
  {D.}~\bibnamefont {Esteve}}, \bibinfo {author} {\bibfnamefont
  {F.}~\bibnamefont {Chiarello}}, \bibinfo {author} {\bibfnamefont
  {A.}~\bibnamefont {Shnirman}}, \bibinfo {author} {\bibfnamefont
  {Y.}~\bibnamefont {Makhlin}}, \bibinfo {author} {\bibfnamefont
  {J.}~\bibnamefont {Schriefl}},\ and\ \bibinfo {author} {\bibfnamefont
  {G.}~\bibnamefont {Sch\"on}},\ }\bibfield  {title} {\bibinfo {title}
  {Decoherence in a superconducting quantum bit circuit},\ }\href
  {https://doi.org/10.1103/PhysRevB.72.134519} {\bibfield  {journal} {\bibinfo
  {journal} {Phys. Rev. B}\ }\textbf {\bibinfo {volume} {72}},\ \bibinfo
  {pages} {134519} (\bibinfo {year} {2005})}\BibitemShut {NoStop}%
\bibitem [{\citenamefont {Janvier}\ \emph {et~al.}(2015)\citenamefont
  {Janvier}, \citenamefont {Tosi}, \citenamefont {Bretheau}, \citenamefont
  {Girit}, \citenamefont {Stern}, \citenamefont {Bertet}, \citenamefont
  {Joyez}, \citenamefont {Vion}, \citenamefont {Esteve}, \citenamefont
  {Goffman}, \citenamefont {Pothier},\ and\ \citenamefont
  {Urbina}}]{janvier2015}%
  \BibitemOpen
  \bibfield  {author} {\bibinfo {author} {\bibfnamefont {C.}~\bibnamefont
  {Janvier}}, \bibinfo {author} {\bibfnamefont {L.}~\bibnamefont {Tosi}},
  \bibinfo {author} {\bibfnamefont {L.}~\bibnamefont {Bretheau}}, \bibinfo
  {author} {\bibfnamefont {{\c{C}}.~{\"{O}}.}\ \bibnamefont {Girit}}, \bibinfo
  {author} {\bibfnamefont {M.}~\bibnamefont {Stern}}, \bibinfo {author}
  {\bibfnamefont {P.}~\bibnamefont {Bertet}}, \bibinfo {author} {\bibfnamefont
  {P.}~\bibnamefont {Joyez}}, \bibinfo {author} {\bibfnamefont
  {D.}~\bibnamefont {Vion}}, \bibinfo {author} {\bibfnamefont {D.}~\bibnamefont
  {Esteve}}, \bibinfo {author} {\bibfnamefont {M.~F.}\ \bibnamefont {Goffman}},
  \bibinfo {author} {\bibfnamefont {H.}~\bibnamefont {Pothier}},\ and\ \bibinfo
  {author} {\bibfnamefont {C.}~\bibnamefont {Urbina}},\ }\bibfield  {title}
  {\bibinfo {title} {Coherent manipulation of andreev states in superconducting
  atomic contacts},\ }\href {https://doi.org/10.1126/science.aab2179}
  {\bibfield  {journal} {\bibinfo  {journal} {Science}\ }\textbf {\bibinfo
  {volume} {349}},\ \bibinfo {pages} {1199} (\bibinfo {year}
  {2015})}\BibitemShut {NoStop}%
\bibitem [{\citenamefont {Hays}\ \emph {et~al.}(2021)\citenamefont {Hays},
  \citenamefont {Fatemi}, \citenamefont {Bouman}, \citenamefont {Cerrillo},
  \citenamefont {Diamond}, \citenamefont {Serniak}, \citenamefont {Connolly},
  \citenamefont {Krogstrup}, \citenamefont {Nygård}, \citenamefont {Yeyati},
  \citenamefont {Geresdi},\ and\ \citenamefont {Devoret}}]{hays2021}%
  \BibitemOpen
  \bibfield  {author} {\bibinfo {author} {\bibfnamefont {M.}~\bibnamefont
  {Hays}}, \bibinfo {author} {\bibfnamefont {V.}~\bibnamefont {Fatemi}},
  \bibinfo {author} {\bibfnamefont {D.}~\bibnamefont {Bouman}}, \bibinfo
  {author} {\bibfnamefont {J.}~\bibnamefont {Cerrillo}}, \bibinfo {author}
  {\bibfnamefont {S.}~\bibnamefont {Diamond}}, \bibinfo {author} {\bibfnamefont
  {K.}~\bibnamefont {Serniak}}, \bibinfo {author} {\bibfnamefont
  {T.}~\bibnamefont {Connolly}}, \bibinfo {author} {\bibfnamefont
  {P.}~\bibnamefont {Krogstrup}}, \bibinfo {author} {\bibfnamefont
  {J.}~\bibnamefont {Nygård}}, \bibinfo {author} {\bibfnamefont {A.~L.}\
  \bibnamefont {Yeyati}}, \bibinfo {author} {\bibfnamefont {A.}~\bibnamefont
  {Geresdi}},\ and\ \bibinfo {author} {\bibfnamefont {M.~H.}\ \bibnamefont
  {Devoret}},\ }\bibfield  {title} {\bibinfo {title} {Coherent manipulation of
  an andreev spin qubit},\ }\href {https://doi.org/10.1126/science.abf0345}
  {\bibfield  {journal} {\bibinfo  {journal} {Science}\ }\textbf {\bibinfo
  {volume} {373}},\ \bibinfo {pages} {430} (\bibinfo {year}
  {2021})}\BibitemShut {NoStop}%
\bibitem [{\citenamefont {Aumentado}\ \emph {et~al.}(2023)\citenamefont
  {Aumentado}, \citenamefont {Catelani},\ and\ \citenamefont
  {Serniak}}]{10.1063/PT.3.5291}%
  \BibitemOpen
  \bibfield  {author} {\bibinfo {author} {\bibfnamefont {J.}~\bibnamefont
  {Aumentado}}, \bibinfo {author} {\bibfnamefont {G.}~\bibnamefont
  {Catelani}},\ and\ \bibinfo {author} {\bibfnamefont {K.}~\bibnamefont
  {Serniak}},\ }\bibfield  {title} {\bibinfo {title} {{Quasiparticle poisoning
  in superconducting quantum computers}},\ }\href
  {https://doi.org/10.1063/PT.3.5291} {\bibfield  {journal} {\bibinfo
  {journal} {Physics Today}\ }\textbf {\bibinfo {volume} {76}},\ \bibinfo
  {pages} {34} (\bibinfo {year} {2023})}\BibitemShut {NoStop}%
\bibitem [{\citenamefont {Joyez}\ \emph {et~al.}(1994)\citenamefont {Joyez},
  \citenamefont {Lafarge}, \citenamefont {Filipe}, \citenamefont {Esteve},\
  and\ \citenamefont {Devoret}}]{PhysRevLett.72.2458}%
  \BibitemOpen
  \bibfield  {author} {\bibinfo {author} {\bibfnamefont {P.}~\bibnamefont
  {Joyez}}, \bibinfo {author} {\bibfnamefont {P.}~\bibnamefont {Lafarge}},
  \bibinfo {author} {\bibfnamefont {A.}~\bibnamefont {Filipe}}, \bibinfo
  {author} {\bibfnamefont {D.}~\bibnamefont {Esteve}},\ and\ \bibinfo {author}
  {\bibfnamefont {M.~H.}\ \bibnamefont {Devoret}},\ }\bibfield  {title}
  {\bibinfo {title} {Observation of parity-induced suppression of josephson
  tunneling in the superconducting single electron transistor},\ }\href
  {https://doi.org/10.1103/PhysRevLett.72.2458} {\bibfield  {journal} {\bibinfo
   {journal} {Phys. Rev. Lett.}\ }\textbf {\bibinfo {volume} {72}},\ \bibinfo
  {pages} {2458} (\bibinfo {year} {1994})}\BibitemShut {NoStop}%
\bibitem [{\citenamefont {Aumentado}\ \emph {et~al.}(2004)\citenamefont
  {Aumentado}, \citenamefont {Keller}, \citenamefont {Martinis},\ and\
  \citenamefont {Devoret}}]{PhysRevLett.92.066802}%
  \BibitemOpen
  \bibfield  {author} {\bibinfo {author} {\bibfnamefont {J.}~\bibnamefont
  {Aumentado}}, \bibinfo {author} {\bibfnamefont {M.~W.}\ \bibnamefont
  {Keller}}, \bibinfo {author} {\bibfnamefont {J.~M.}\ \bibnamefont
  {Martinis}},\ and\ \bibinfo {author} {\bibfnamefont {M.~H.}\ \bibnamefont
  {Devoret}},\ }\bibfield  {title} {\bibinfo {title} {Nonequilibrium
  quasiparticles and $2e$ periodicity in single-cooper-pair transistors},\
  }\href {https://doi.org/10.1103/PhysRevLett.92.066802} {\bibfield  {journal}
  {\bibinfo  {journal} {Phys. Rev. Lett.}\ }\textbf {\bibinfo {volume} {92}},\
  \bibinfo {pages} {066802} (\bibinfo {year} {2004})}\BibitemShut {NoStop}%
\bibitem [{\citenamefont {Martinis}\ \emph {et~al.}(2009)\citenamefont
  {Martinis}, \citenamefont {Ansmann},\ and\ \citenamefont
  {Aumentado}}]{PhysRevLett.103.097002}%
  \BibitemOpen
  \bibfield  {author} {\bibinfo {author} {\bibfnamefont {J.~M.}\ \bibnamefont
  {Martinis}}, \bibinfo {author} {\bibfnamefont {M.}~\bibnamefont {Ansmann}},\
  and\ \bibinfo {author} {\bibfnamefont {J.}~\bibnamefont {Aumentado}},\
  }\bibfield  {title} {\bibinfo {title} {Energy decay in superconducting
  josephson-junction qubits from nonequilibrium quasiparticle excitations},\
  }\href {https://doi.org/10.1103/PhysRevLett.103.097002} {\bibfield  {journal}
  {\bibinfo  {journal} {Phys. Rev. Lett.}\ }\textbf {\bibinfo {volume} {103}},\
  \bibinfo {pages} {097002} (\bibinfo {year} {2009})}\BibitemShut {NoStop}%
\bibitem [{\citenamefont {Huang}\ \emph {et~al.}(2026)\citenamefont {Huang},
  \citenamefont {Ortuzar},\ and\ \citenamefont {Cazalilla}}]{zenodo}%
  \BibitemOpen
  \bibfield  {author} {\bibinfo {author} {\bibfnamefont {C.-H.}\ \bibnamefont
  {Huang}}, \bibinfo {author} {\bibfnamefont {J.}~\bibnamefont {Ortuzar}},\
  and\ \bibinfo {author} {\bibfnamefont {M.~A.}\ \bibnamefont {Cazalilla}},\
  }\bibfield  {title} {\bibinfo {title} {Datas for publication, superconducting
  spin- singlet qubit in a triangulene spin chain},\ }\href
  {https://doi.org/10.5281/zenodo.18146812} {10.5281/zenodo.18146812} (\bibinfo
  {year} {2026})\BibitemShut {NoStop}%
\bibitem [{\citenamefont {Ortuzar}\ \emph {et~al.}(2023)\citenamefont
  {Ortuzar}, \citenamefont {Pascual}, \citenamefont {Bergeret},\ and\
  \citenamefont {Cazalilla}}]{ortuzar2023}%
  \BibitemOpen
  \bibfield  {author} {\bibinfo {author} {\bibfnamefont {J.}~\bibnamefont
  {Ortuzar}}, \bibinfo {author} {\bibfnamefont {J.~I.}\ \bibnamefont
  {Pascual}}, \bibinfo {author} {\bibfnamefont {F.~S.}\ \bibnamefont
  {Bergeret}},\ and\ \bibinfo {author} {\bibfnamefont {M.~A.}\ \bibnamefont
  {Cazalilla}},\ }\bibfield  {title} {\bibinfo {title} {Theory of a single
  magnetic impurity on a thin metal film in proximity to a superconductor},\
  }\href {https://doi.org/10.1103/PhysRevB.108.024511} {\bibfield  {journal}
  {\bibinfo  {journal} {Phys. Rev. B}\ }\textbf {\bibinfo {volume} {108}},\
  \bibinfo {pages} {024511} (\bibinfo {year} {2023})}\BibitemShut {NoStop}%
\bibitem [{\citenamefont {Gyenis}\ \emph
  {et~al.}(2021{\natexlab{a}})\citenamefont {Gyenis}, \citenamefont {Di~Paolo},
  \citenamefont {Koch}, \citenamefont {Blais}, \citenamefont {Houck},\ and\
  \citenamefont {Schuster}}]{PRXQuantum.2.030101}%
  \BibitemOpen
  \bibfield  {author} {\bibinfo {author} {\bibfnamefont {A.}~\bibnamefont
  {Gyenis}}, \bibinfo {author} {\bibfnamefont {A.}~\bibnamefont {Di~Paolo}},
  \bibinfo {author} {\bibfnamefont {J.}~\bibnamefont {Koch}}, \bibinfo {author}
  {\bibfnamefont {A.}~\bibnamefont {Blais}}, \bibinfo {author} {\bibfnamefont
  {A.~A.}\ \bibnamefont {Houck}},\ and\ \bibinfo {author} {\bibfnamefont
  {D.~I.}\ \bibnamefont {Schuster}},\ }\bibfield  {title} {\bibinfo {title}
  {Moving beyond the transmon: Noise-protected superconducting quantum
  circuits},\ }\href {https://doi.org/10.1103/PRXQuantum.2.030101} {\bibfield
  {journal} {\bibinfo  {journal} {PRX Quantum}\ }\textbf {\bibinfo {volume}
  {2}},\ \bibinfo {pages} {030101} (\bibinfo {year}
  {2021}{\natexlab{a}})}\BibitemShut {NoStop}%
\bibitem [{\citenamefont {Smith}\ \emph {et~al.}()\citenamefont {Smith},
  \citenamefont {Kou}, \citenamefont {Xiao}, \citenamefont {Vool},\ and\
  \citenamefont {Devoret}}]{smith_superconducting_2020}%
  \BibitemOpen
  \bibfield  {author} {\bibinfo {author} {\bibfnamefont {W.~C.}\ \bibnamefont
  {Smith}}, \bibinfo {author} {\bibfnamefont {A.}~\bibnamefont {Kou}}, \bibinfo
  {author} {\bibfnamefont {X.}~\bibnamefont {Xiao}}, \bibinfo {author}
  {\bibfnamefont {U.}~\bibnamefont {Vool}},\ and\ \bibinfo {author}
  {\bibfnamefont {M.~H.}\ \bibnamefont {Devoret}},\ }\bibfield  {title}
  {\bibinfo {title} {Superconducting circuit protected by two-cooper-pair
  tunneling},\ }\href {https://doi.org/10.1038/s41534-019-0231-2} {\ \textbf
  {\bibinfo {volume} {6}},\ \bibinfo {pages} {8}}\BibitemShut {NoStop}%
\bibitem [{\citenamefont {Gyenis}\ \emph
  {et~al.}(2021{\natexlab{b}})\citenamefont {Gyenis}, \citenamefont {Di~Paolo},
  \citenamefont {Koch}, \citenamefont {Blais}, \citenamefont {Houck},\ and\
  \citenamefont {Schuster}}]{Gyenis2021}%
  \BibitemOpen
  \bibfield  {author} {\bibinfo {author} {\bibfnamefont {A.}~\bibnamefont
  {Gyenis}}, \bibinfo {author} {\bibfnamefont {A.}~\bibnamefont {Di~Paolo}},
  \bibinfo {author} {\bibfnamefont {J.}~\bibnamefont {Koch}}, \bibinfo {author}
  {\bibfnamefont {A.}~\bibnamefont {Blais}}, \bibinfo {author} {\bibfnamefont
  {A.~A.}\ \bibnamefont {Houck}},\ and\ \bibinfo {author} {\bibfnamefont
  {D.~I.}\ \bibnamefont {Schuster}},\ }\bibfield  {title} {\bibinfo {title}
  {Moving beyond the transmon: Noise-protected superconducting quantum
  circuits},\ }\href {https://doi.org/10.1103/PRXQuantum.2.030101} {\bibfield
  {journal} {\bibinfo  {journal} {PRX Quantum}\ }\textbf {\bibinfo {volume}
  {2}},\ \bibinfo {pages} {030101} (\bibinfo {year}
  {2021}{\natexlab{b}})}\BibitemShut {NoStop}%
\bibitem [{\citenamefont {{\v Z}itko}(2009)}]{ZITKO20091271_discret}%
  \BibitemOpen
  \bibfield  {author} {\bibinfo {author} {\bibfnamefont {R.}~\bibnamefont {{\v
  Z}itko}},\ }\bibfield  {title} {\bibinfo {title} {Adaptive logarithmic
  discretization for numerical renormalization group methods},\ }\href
  {https://doi.org/https://doi.org/10.1016/j.cpc.2009.02.007} {\bibfield
  {journal} {\bibinfo  {journal} {Computer Physics Communications}\ }\textbf
  {\bibinfo {volume} {180}},\ \bibinfo {pages} {1271} (\bibinfo {year}
  {2009})}\BibitemShut {NoStop}%
\bibitem [{\citenamefont {Yao}\ \emph {et~al.}(2014{\natexlab{b}})\citenamefont
  {Yao}, \citenamefont {Moca}, \citenamefont {Weymann}, \citenamefont {Sau},
  \citenamefont {Lukin}, \citenamefont {Demler},\ and\ \citenamefont
  {Zar\'and}}]{PhysRevB.90.241108_YAO}%
  \BibitemOpen
  \bibfield  {author} {\bibinfo {author} {\bibfnamefont {N.~Y.}\ \bibnamefont
  {Yao}}, \bibinfo {author} {\bibfnamefont {C.~P.}\ \bibnamefont {Moca}},
  \bibinfo {author} {\bibfnamefont {I.}~\bibnamefont {Weymann}}, \bibinfo
  {author} {\bibfnamefont {J.~D.}\ \bibnamefont {Sau}}, \bibinfo {author}
  {\bibfnamefont {M.~D.}\ \bibnamefont {Lukin}}, \bibinfo {author}
  {\bibfnamefont {E.~A.}\ \bibnamefont {Demler}},\ and\ \bibinfo {author}
  {\bibfnamefont {G.}~\bibnamefont {Zar\'and}},\ }\bibfield  {title} {\bibinfo
  {title} {Phase diagram and excitations of a shiba molecule},\ }\href
  {https://doi.org/10.1103/PhysRevB.90.241108} {\bibfield  {journal} {\bibinfo
  {journal} {Phys. Rev. B}\ }\textbf {\bibinfo {volume} {90}},\ \bibinfo
  {pages} {241108} (\bibinfo {year} {2014}{\natexlab{b}})}\BibitemShut
  {NoStop}%
\bibitem [{\citenamefont {T\'oth}\ \emph {et~al.}(2008)\citenamefont {T\'oth},
  \citenamefont {Moca}, \citenamefont {Legeza},\ and\ \citenamefont
  {Zar\'and}}]{PhysRevB.78.245109_symmetry}%
  \BibitemOpen
  \bibfield  {author} {\bibinfo {author} {\bibfnamefont {A.~I.}\ \bibnamefont
  {T\'oth}}, \bibinfo {author} {\bibfnamefont {C.~P.}\ \bibnamefont {Moca}},
  \bibinfo {author} {\bibfnamefont {O.}~\bibnamefont {Legeza}},\ and\ \bibinfo
  {author} {\bibfnamefont {G.}~\bibnamefont {Zar\'and}},\ }\bibfield  {title}
  {\bibinfo {title} {Density matrix numerical renormalization group for
  non-abelian symmetries},\ }\href {https://doi.org/10.1103/PhysRevB.78.245109}
  {\bibfield  {journal} {\bibinfo  {journal} {Phys. Rev. B}\ }\textbf {\bibinfo
  {volume} {78}},\ \bibinfo {pages} {245109} (\bibinfo {year}
  {2008})}\BibitemShut {NoStop}%
\bibitem [{\citenamefont {Satori}\ \emph {et~al.}(1992)\citenamefont {Satori},
  \citenamefont {Shiba}, \citenamefont {Sakai},\ and\ \citenamefont
  {Shimizu}}]{JPSJ.67.1332_BCS_trans}%
  \BibitemOpen
  \bibfield  {author} {\bibinfo {author} {\bibfnamefont {K.}~\bibnamefont
  {Satori}}, \bibinfo {author} {\bibfnamefont {H.}~\bibnamefont {Shiba}},
  \bibinfo {author} {\bibfnamefont {O.}~\bibnamefont {Sakai}},\ and\ \bibinfo
  {author} {\bibfnamefont {Y.}~\bibnamefont {Shimizu}},\ }\bibfield  {title}
  {\bibinfo {title} {Numerical renormalization group study of magnetic
  impurities in superconductors},\ }\href
  {https://doi.org/10.1143/JPSJ.61.3239} {\bibfield  {journal} {\bibinfo
  {journal} {Journal of the Physical Society of Japan}\ }\textbf {\bibinfo
  {volume} {61}},\ \bibinfo {pages} {3239} (\bibinfo {year}
  {1992})}\BibitemShut {NoStop}%
\bibitem [{\citenamefont {Hecht}\ \emph {et~al.}(2008)\citenamefont {Hecht},
  \citenamefont {Weichselbaum}, \citenamefont {von Delft},\ and\ \citenamefont
  {Bulla}}]{Hecht_2008_BCS}%
  \BibitemOpen
  \bibfield  {author} {\bibinfo {author} {\bibfnamefont {T.}~\bibnamefont
  {Hecht}}, \bibinfo {author} {\bibfnamefont {A.}~\bibnamefont {Weichselbaum}},
  \bibinfo {author} {\bibfnamefont {J.}~\bibnamefont {von Delft}},\ and\
  \bibinfo {author} {\bibfnamefont {R.}~\bibnamefont {Bulla}},\ }\bibfield
  {title} {\bibinfo {title} {Numerical renormalization group calculation of
  near-gap peaks in spectral functions of the anderson model with
  superconducting leads},\ }\href
  {https://doi.org/10.1088/0953-8984/20/27/275213} {\bibfield  {journal}
  {\bibinfo  {journal} {Journal of Physics: Condensed Matter}\ }\textbf
  {\bibinfo {volume} {20}},\ \bibinfo {pages} {275213} (\bibinfo {year}
  {2008})}\BibitemShut {NoStop}%
\bibitem [{\citenamefont {Weichselbaum}\ and\ \citenamefont {von
  Delft}(2007)}]{PhysRevLett.99.076402_FDM}%
  \BibitemOpen
  \bibfield  {author} {\bibinfo {author} {\bibfnamefont {A.}~\bibnamefont
  {Weichselbaum}}\ and\ \bibinfo {author} {\bibfnamefont {J.}~\bibnamefont {von
  Delft}},\ }\bibfield  {title} {\bibinfo {title} {Sum-rule conserving spectral
  functions from the numerical renormalization group},\ }\href
  {https://doi.org/10.1103/PhysRevLett.99.076402} {\bibfield  {journal}
  {\bibinfo  {journal} {Phys. Rev. Lett.}\ }\textbf {\bibinfo {volume} {99}},\
  \bibinfo {pages} {076402} (\bibinfo {year} {2007})}\BibitemShut {NoStop}%
\bibitem [{\citenamefont {Anders}\ and\ \citenamefont
  {Schiller}(2005)}]{AS_PhysRevLett.95.196801}%
  \BibitemOpen
  \bibfield  {author} {\bibinfo {author} {\bibfnamefont {F.~B.}\ \bibnamefont
  {Anders}}\ and\ \bibinfo {author} {\bibfnamefont {A.}~\bibnamefont
  {Schiller}},\ }\bibfield  {title} {\bibinfo {title} {Real-time dynamics in
  quantum-impurity systems: A time-dependent numerical renormalization-group
  approach},\ }\href {https://doi.org/10.1103/PhysRevLett.95.196801} {\bibfield
   {journal} {\bibinfo  {journal} {Phys. Rev. Lett.}\ }\textbf {\bibinfo
  {volume} {95}},\ \bibinfo {pages} {196801} (\bibinfo {year}
  {2005})}\BibitemShut {NoStop}%
\bibitem [{\citenamefont {Anders}\ and\ \citenamefont
  {Schiller}(2006)}]{td_NRG_Smatrix_PhysRevB.74.245113}%
  \BibitemOpen
  \bibfield  {author} {\bibinfo {author} {\bibfnamefont {F.~B.}\ \bibnamefont
  {Anders}}\ and\ \bibinfo {author} {\bibfnamefont {A.}~\bibnamefont
  {Schiller}},\ }\bibfield  {title} {\bibinfo {title} {Spin precession and
  real-time dynamics in the kondo model: Time-dependent numerical
  renormalization-group study},\ }\href
  {https://doi.org/10.1103/PhysRevB.74.245113} {\bibfield  {journal} {\bibinfo
  {journal} {Phys. Rev. B}\ }\textbf {\bibinfo {volume} {74}},\ \bibinfo
  {pages} {245113} (\bibinfo {year} {2006})}\BibitemShut {NoStop}%
\bibitem [{\citenamefont {Weymann}\ \emph {et~al.}(2015)\citenamefont
  {Weymann}, \citenamefont {von Delft},\ and\ \citenamefont
  {Weichselbaum}}]{FDM_identity_PhysRevB.92.155435}%
  \BibitemOpen
  \bibfield  {author} {\bibinfo {author} {\bibfnamefont {I.}~\bibnamefont
  {Weymann}}, \bibinfo {author} {\bibfnamefont {J.}~\bibnamefont {von Delft}},\
  and\ \bibinfo {author} {\bibfnamefont {A.}~\bibnamefont {Weichselbaum}},\
  }\bibfield  {title} {\bibinfo {title} {Thermalization and dynamics in the
  single-impurity anderson model},\ }\href
  {https://doi.org/10.1103/PhysRevB.92.155435} {\bibfield  {journal} {\bibinfo
  {journal} {Phys. Rev. B}\ }\textbf {\bibinfo {volume} {92}},\ \bibinfo
  {pages} {155435} (\bibinfo {year} {2015})}\BibitemShut {NoStop}%
\bibitem [{\citenamefont {Yao}\ \emph {et~al.}(2014{\natexlab{c}})\citenamefont
  {Yao}, \citenamefont {Glazman}, \citenamefont {Demler}, \citenamefont
  {Lukin},\ and\ \citenamefont {Sau}}]{PhysRevLett.113.087202RKKY}%
  \BibitemOpen
  \bibfield  {author} {\bibinfo {author} {\bibfnamefont {N.~Y.}\ \bibnamefont
  {Yao}}, \bibinfo {author} {\bibfnamefont {L.~I.}\ \bibnamefont {Glazman}},
  \bibinfo {author} {\bibfnamefont {E.~A.}\ \bibnamefont {Demler}}, \bibinfo
  {author} {\bibfnamefont {M.~D.}\ \bibnamefont {Lukin}},\ and\ \bibinfo
  {author} {\bibfnamefont {J.~D.}\ \bibnamefont {Sau}},\ }\bibfield  {title}
  {\bibinfo {title} {Enhanced antiferromagnetic exchange between magnetic
  impurities in a superconducting host},\ }\href
  {https://doi.org/10.1103/PhysRevLett.113.087202} {\bibfield  {journal}
  {\bibinfo  {journal} {Phys. Rev. Lett.}\ }\textbf {\bibinfo {volume} {113}},\
  \bibinfo {pages} {087202} (\bibinfo {year} {2014}{\natexlab{c}})}\BibitemShut
  {NoStop}%
\bibitem [{\citenamefont {Baumann}\ \emph {et~al.}(2015)\citenamefont
  {Baumann}, \citenamefont {Paul}, \citenamefont {Choi}, \citenamefont {Lutz},
  \citenamefont {Ardavan},\ and\ \citenamefont {Heinrich}}]{Baumann2015}%
  \BibitemOpen
  \bibfield  {author} {\bibinfo {author} {\bibfnamefont {S.}~\bibnamefont
  {Baumann}}, \bibinfo {author} {\bibfnamefont {W.}~\bibnamefont {Paul}},
  \bibinfo {author} {\bibfnamefont {T.}~\bibnamefont {Choi}}, \bibinfo {author}
  {\bibfnamefont {C.~P.}\ \bibnamefont {Lutz}}, \bibinfo {author}
  {\bibfnamefont {A.}~\bibnamefont {Ardavan}},\ and\ \bibinfo {author}
  {\bibfnamefont {A.~J.}\ \bibnamefont {Heinrich}},\ }\bibfield  {title}
  {\bibinfo {title} {Electron paramagnetic resonance of individual atoms on a
  surface},\ }\href {https://doi.org/10.1126/science.aac8703} {\bibfield
  {journal} {\bibinfo  {journal} {Science}\ }\textbf {\bibinfo {volume}
  {350}},\ \bibinfo {pages} {417} (\bibinfo {year} {2015})}\BibitemShut
  {NoStop}%
\bibitem [{\citenamefont {Sellies}\ \emph {et~al.}(2023)\citenamefont
  {Sellies}, \citenamefont {Spachtholz}, \citenamefont {Bleher}, \citenamefont
  {Eckrich}, \citenamefont {Scheuerer},\ and\ \citenamefont
  {Repp}}]{Sellies2023}%
  \BibitemOpen
  \bibfield  {author} {\bibinfo {author} {\bibfnamefont {L.}~\bibnamefont
  {Sellies}}, \bibinfo {author} {\bibfnamefont {R.}~\bibnamefont {Spachtholz}},
  \bibinfo {author} {\bibfnamefont {S.}~\bibnamefont {Bleher}}, \bibinfo
  {author} {\bibfnamefont {J.}~\bibnamefont {Eckrich}}, \bibinfo {author}
  {\bibfnamefont {P.}~\bibnamefont {Scheuerer}},\ and\ \bibinfo {author}
  {\bibfnamefont {J.}~\bibnamefont {Repp}},\ }\bibfield  {title} {\bibinfo
  {title} {Single-molecule electron spin resonance by means of atomic force
  microscopy},\ }\href {https://doi.org/10.1038/s41586-023-06754-6} {\bibfield
  {journal} {\bibinfo  {journal} {Nature}\ }\textbf {\bibinfo {volume} {624}},\
  \bibinfo {pages} {64} (\bibinfo {year} {2023})}\BibitemShut {NoStop}%
\bibitem [{\citenamefont {Huang}\ \emph {et~al.}(2025)\citenamefont {Huang},
  \citenamefont {Ortuzar},\ and\ \citenamefont {Cazalilla}}]{unpub2025}%
  \BibitemOpen
  \bibfield  {author} {\bibinfo {author} {\bibfnamefont {C.-H.}\ \bibnamefont
  {Huang}}, \bibinfo {author} {\bibfnamefont {J.}~\bibnamefont {Ortuzar}},\
  and\ \bibinfo {author} {\bibfnamefont {M.~A.}\ \bibnamefont {Cazalilla}},\
  }\bibfield  {title} {\bibinfo {title} {To be published},\ }\href@noop {} {\
  (\bibinfo {year} {2025})},\ \bibinfo {note} {(unpublished)}\BibitemShut
  {NoStop}%
\end{thebibliography}%

\end{document}